\documentclass[10pt,a4paper]{article}

\usepackage{epsfig}
\usepackage{graphicx}
\usepackage{tabularx}
\usepackage{amsmath}
\usepackage{amssymb}
\usepackage{amstext}
\usepackage{amsfonts}
\usepackage[round]{natbib}

\newcommand{\kms}[0]{\unskip\ensuremath{\,\textup{km\,s}^{-1}}}
\newcommand{\cm}[1]{\unskip\ensuremath{\,{\rm cm}^{#1}}}
\newcommand{\bm}[1]{\ensuremath\mathbf{#1}}
\newcommand{\avrg}[1]{\ensuremath{\langle #1 \rangle}}
\newcommand{\ch}[1]{\mbox{$\mathrm{#1}$}}
\newcommand{\molt}[0]{MT}
\newcommand{\isot}[0]{IT}

\title{Hydrodynamical simulations of the decay of high-speed
molecular turbulence. II. Divergence from isothermality }

\author{Georgi Pavlovski,$^1$%
\footnote{Email: gbp@phys.soton.ac.uk (GBP); m.d.smith@kent.ac.uk (MDS); mordecai@amnh.org (M-MML)},
Michael D. Smith,$^{2,3}$\\
and Mordecai-Mark Mac Low~$^4$\\
~
\\
\parbox[b]{0.85\linewidth}{\tiny$^1$ School of Physics and Astronomy, University of Southampton,
Southampton SO17 1BJ, U.K.\\
$^2$ Armagh Observatory, College Hill, Armagh BT61 9DG,
Northern Ireland, U.K. \\
$^3$ Centre for Astrophysics \& Planetary Science, 
  The University of Kent, Canterbury CT2 7NR, U.K.\\ 
$^4$ Department of Astrophysics, American Museum of Natural History,
79th Street at Central Park West, New York, NY 10024-5192, USA}}

\date{\today}

\begin{document}

\maketitle

\begin{abstract}
A roughly constant temperature over a wide range of densities is maintained in molecular
clouds through radiative heating and cooling. An isothermal equation of state is therefore 
frequently employed in molecular cloud simulations. However, the dynamical processes in 
molecular clouds include shock waves, expansion waves, cooling induced collapse and 
baroclinic vorticity, all incompatible with the assumption of a purely isothermal flow.
Here, we incorporate an energy equation including all the important heating and cooling rates
and a simple chemical network into simulations of three-dimensional, hydrodynamic, 
decaying turbulence. This allows us to test the accuracy of the isothermal assumption 
by directly comparing a model run with the modified energy equation to an isothermal model. 
We compute an extreme case in which the initial turbulence is sufficiently strong to
dissociate much of the gas and alter the specific heat ratio. The molecules then reform 
as the turbulence weakens. We track the true specific heat ratio as well as its effective 
value. We analyse power spectra, vorticity and shock structures, and discuss scaling 
laws for decaying turbulence. We derive some limitations to the isothermal approximation 
for simulations of the interstellar medium using simple projection techniques. Overall, 
even given the extreme conditions, we find that an isothermal flow provides 
an adequate physical and observational description of many properties. The main exceptions 
revealed here concern behaviour directly related to the high temperature zones 
behind the shock waves. 

\end{abstract}


\section{INTRODUCTION} 
\label{sec:int}

State-of-the-art numerical simulations of star forming regions take into account 
a small selection from a wide range of relevant physical processes. Of these, it 
has been found that supersonic turbulent motions combined with self-gravity and 
magnetic field reproduce observations of the morphology and 
dynamics of molecular clouds on all scales
(\citealt{Tilley04,Bonnell04,Bonnell03,Padoan04,Padoan03};  
\citealt*[see][for reviews]{Elmegreen04,M-MML03}).   
However, in order to reduce the computational constraints, many results are based
on simulations of an isothermal gas. This implies (1) highly radiative shocks, 
(2) suppression of the steepening of sound waves, (3) suppression of thermal 
instability and shock wave overstability, (4) no adiabatic cooling in expanding 
regions, (5) suppression of baroclinic vorticity and (6) helicity conservation. 
These assumptions are not always valid and could lead to erroneous results. 
In addition, one can examine neither the influence of the internal properties of 
the gas, such as its chemical composition, on the turbulent fluid flow nor the 
effect of the field of shock waves on the chemical composition.

As one approach toward a more realistic scenario, we have performed numerical 
simulations of supersonic turbulence in dense, initially uniform, molecular material, 
with an equation of state determined by the chemical composition of the gas 
\citep[][; hereafter Paper I]{gbp02}. We include a small but relevant chemical reaction
network, based on the work of \citet{Smith03}. We follow the time-dependent hydrogen
chemistry, with \ch{C} and \ch{O} chemistry limited to reactions with neutral atomic  
and molecular hydrogen, which generated \ch{OH}, \ch{CO} and \ch{H_2O} 
\citep[see][Appendix~B]{Smith03}.  Our cooling function contains \ch{H_2} 
ro-vibrational and dissociative cooling, \ch{CO} and \ch{H_2O} ro-vibrational
cooling, gas-grain, thermal bremsstrahlung and a steady-state approximation
to atomic cooling. That is, we follow shock-enhanced chemistry rather than the 
chemistry of the cold molecular gas.

We demonstrated in Paper~I, quite surprisingly, that even in the somewhat extreme 
case of high speed \ch{H_2} dissociative turbulence, many integrated properties of 
turbulence do not differ much from the corresponding properties in the isothermal regime.
In particular, the rate of loss of energy in decaying turbulence obeys a similar law.
Here, we examine non-isothermal turbulence in a broader context to better understand 
where to trust the isothermal approximation.

Following the chemistry reveals properties not encountered in the isothermal 
simulations. Most importantly, we found that supersonic turbulence speeds up 
the reformation of hydrogen molecules by a factor of a few. This was
developed by \citet{Smith02} into a scenario for simultaneous rapid molecule and rapid
molecular cloud formation out of atomic clouds. The accelerated chemistry 
occurs in the dense layers behind shock waves where atomic collisional rates are enhanced. 
After a dynamical time, the reformed molecules are widely distributed as most of the 
gas has at some time passed through a shock.

Here, we examine three dimensional decaying supersonic turbulence. We restrict our 
analysis to the high speed case of Paper~I (see \S~\ref{sec:setup} for description of the initial 
conditions; turbulence is chosen to be initially strong enough to destroy the molecules). 
The molecules subsequently 
reform and the field of shock waves then slowly decays within the molecular gas.
Hereafter, we will use molecular turbulence (\molt) as a shorthand for the run using
the molecular chemistry network data and isothermal turbulence (\isot) for the run using 
the isothermal equation of state.

In \S~\ref{sec:res}, the power spectrum and cascade are analysed. Probability distribution 
functions (PDFs) for density, molecular density and velocity are presented in \S~\ref{sec:vsh}.
The observational implications are considered in \S~\ref{sec:obs}, Over the years, a
number of different techniques have been proposed to interface analyses of numerical simulations 
with the observations \citep{Brunt03,Heyer97,Ossenkopf02,M-MML00,Padoan98}.  
To perform our comparison study, we apply here the same techniques to simulated 
observations derived from models based on both the chemical network described above and the 
isothermal assumption.

\section{The Simulations}
\label{sec:setup}

The gas is modelled within 
a three-dimensional 
box with periodic boundary conditions,
simulated on a grid of 256$^3$ zones
using a modified version of the gas dynamics code ZEUS-3D \citep{Stone92I}.
Paper~I gives full details on the modifications to ZEUS-3D and on our
initial set up.

The initial state of the gas was 
chosen so that we could
investigate a diverse range of conditions in the subsequent evolution.
The gas is initially fully molecular. The imposed velocity field corresponds to that of 
the highest speed turbulence from Paper~I, with a root-mean-square 
velocity of $v_{\text{rms}}$ = (60\,\kms)$v_{60}$. The turbulence is allowed to decay. In this manner,
we can study the initial brief period within which shocks form
and molecules dissociate. This is of order of the dynamical time

\begin{equation}
   t_{\text{dyn}} = L/(k v_{\text{rms}}) = (15\, {\rm yr}) L_{16} v_{60}^{-1} (k/3.5)^{-1},
\end{equation}
for a box of size $L = (10^{16}$~cm)$L_{16}$ and a mean 
driving wavenumber $k$.
This is followed by a period of molecule reformation of expected duration
\begin{equation}
   t_{\text{ref}} = 10^{10}/(n T^{1/2})\,  {\rm yr} = (1060\, {\rm yr})n_6^{-1} T_{100}^{-1/2},
\end{equation}
for a hydrogen nucleon density of $n = (10^6$ \cm{-3})$n_6$ and
initial temperature $T=(100 \mbox{ K})T_{100}$.
This stage is considerably shortened by the turbulent compression.  Finally, 
we follow an extended period of gradually decaying molecular
turbulence.  
We set our numerical initial conditions to have $L_{16} = n_6 = v_{60} = T_{100} =
1$.

It should be noted that the total cooling length behind a 
fast dissociative shock at the above density is $\sim$~10$^{13}$\,cm 
\citep{Smith03} and is not resolved in the simulation,
so we cannot  predict quantities or 
instabilities inherent to the shock physics and dynamics.
The corresponding cooling time is approximately
\begin{equation}
   t_{\text{cool}} = (1 \mbox{ yr}) n_6^{-1}.
\end{equation}
However, mass and momentum are still conserved, equilibrium chemistry is maintained, and the
dissociation speed limits are approximately correct (Paper~I). In the subsequent slower shocks
within the molecular turbulence, the radiative layers are thicker and partly resolved.

The temperature in \isot\ was fixed at 100\,K. All other parameters were unchanged from the
\molt\ case. That includes the initial kinetic energy spectrum which is imposed  artificially
by perturbing the velocity field with a Gaussian random field following \citet{M-MML98}. 
A narrow range of wavenumbers, $3<|\bm{k}|<4$, is chosen for the velocity perturbation.

It should be noted that the molecular simulation still excludes certain aspects of a molecular supersonic
flow.  The resolution precludes the appearance of the shock overstability \citep{Smith03}. 
The zone size limits the 
degree of compression although high densities can occur when material accumulates within a zone.
Both these effects can only be adequately modelled by decreasing the zone size by a 
factor of 100 or by introducing a strong magnetic field.  Equilibrium chemistry, however, allows us to 
proceed and to follow some of the consequences of non-isothermal shock physics.

\section{POWER SPECTRA}
\label{sec:eng}

\subsection{Definitions}
\label{sec:def}

Power spectra of the fluid variables have long been used to characterise turbulence and 
so provide a standard means of comparing simulations. The total spectral power density is
\begin{equation}
\label{eng:eq3}
  P(k) = \sum_{k<|\bm{k}|<k+1} |\hat{\bm{v}}(\bm{k})|^2,
\end{equation}
where $\hat{\bm{v}}(\bm{k})=\cal{F}\left[\bm{v}\left(\bm{r}\right)\right]$ is
the Fourier transform of the velocity field (and $\cal{F}$ is a Fourier operator).

According to the Helmholtz decomposition theorem, any differentiable vector field with 
divergence $\left(\nabla \cdot \bm{v}\right)$ and curl $\left[\nabla\times\bm{v}\right]$ can
be split into compressional and solenoidal components. We thus write the velocity of the fluid as
\begin{equation}
\label{eng:eq1}
  \bm{v}\left(\bm{r}\right) = \bm{v}_s\left(\bm{r}\right) + 
  \bm{v}_c\left(\bm{r}\right),
\end{equation}
where
\begin{equation}
\label{eng:eq2}
  \left[\nabla\times\bm{v}_c\right] = 0,\qquad
  \left(\nabla.\bm{v}_s\right) = 0.
\end{equation}
The solenoidal and compressional power spectra are written as
\begin{equation}
\label{eng:eq4}
  P_s(k) = \sum_{k<|\bm{k}|<k+1} |\hat{\bm{v}}_s(\bm{k})|^2
\end{equation}
and
\begin{equation}
\label{eng:eq5}
  P_c(k) = \sum_{k<|\bm{k}|<k+1} |\hat{\bm{v}}_c(\bm{k})|^2.
\end{equation}
The total power spectrum can also be expressed as $P(k) = P_s(k) + P_c(k)$ since
$\left(\hat{\bm{v}}_s^\ast.\hat{\bm{v}}_c\right)=0$. Note that we consider here angle averaged
quantities, which is appropriate for isotropic turbulence (by isotropic we mean that the correlation 
length is smaller than the computational domain; see Paper~I for details).

Initially, the power spectrum of velocity also corresponds
to the kinetic energy distribution (or `energy spectrum') since the density is 
initially uniform.  The initial energy is divided one-third into
compressive modes and two-thirds into solenoidal modes.

\subsection{Total Power Spectra}
\label{sec:res}

\begin{figure}
\centering\includegraphics{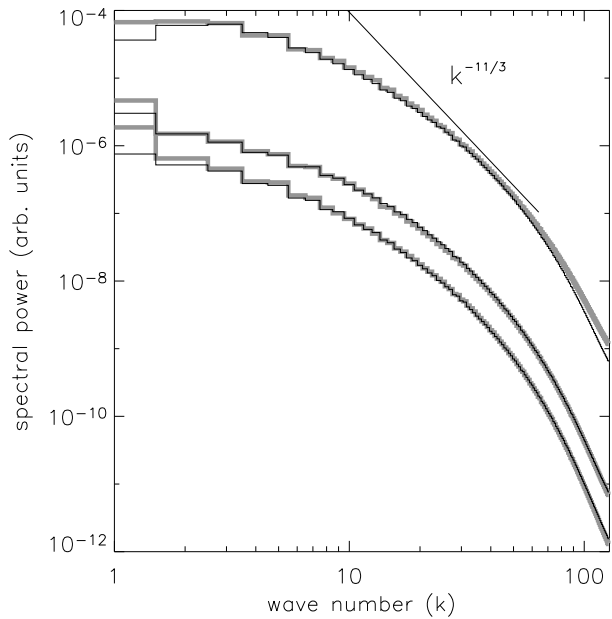}
\caption[Velocity power spectrum evolution]%
{Evolution of the velocity power spectrum for \isot\ case (thick grey lines) and 
\molt\ (thin black lines) simulations. Data at three instances are displayed: upper lines: $t=30$~yr; 
medium lines: $t=300$~yr; lower lines: $t=600$~yr.  See also Fig.~\ref{eng:fig2a}.}
\label{eng:fig2}
\end{figure}

The spectral peak rapidly widens at the beginning of the simulations, as illustrated by the 
curves in Fig.~\ref{eng:fig2} for \molt\ (thin black line) and \isot\ (thick grey line). 
It develops into a wavenumber cascade with a maximum in the same initial range 
$3<|\bm{k}|<4$ for \molt. The minimum is located at the high $k$ end. 

At later times, the power is redistributed into a monotonic cascade
with the maximum at $k = 1$ and minimum at $k = 127$  (see Figs.~\ref{eng:fig2}). 
For comparison, we indicate on the diagram the power-law Kolmogorov spectrum, $k^{-11/3}$, 
which is expected to occur even in {\em decaying} three dimensional incompressible turbulence 
\citep{Lesieur97}. In contrast,
no clear power law develops in the spectra here as there is no continuing
input of energy in the driving range in these decaying simulations {\em and} the power falls 
somewhat faster  at high wavenumbers than at low wavenumbers in both cases (in this respect, 
the apparent 'convergence' of the curves at the high $k$ end is illusory; see Table~\ref{eng:tab1} 
for the quantitative measure of the slope angle). In terms of the spectrum of
shock waves, the increasing dominance of weak shocks preferentially dissipates the energy
on small scales (see \S~\ref{ssec:vels}).

\begin{figure}
\begin{minipage}[b]{.49\linewidth}
\includegraphics[width=\linewidth]{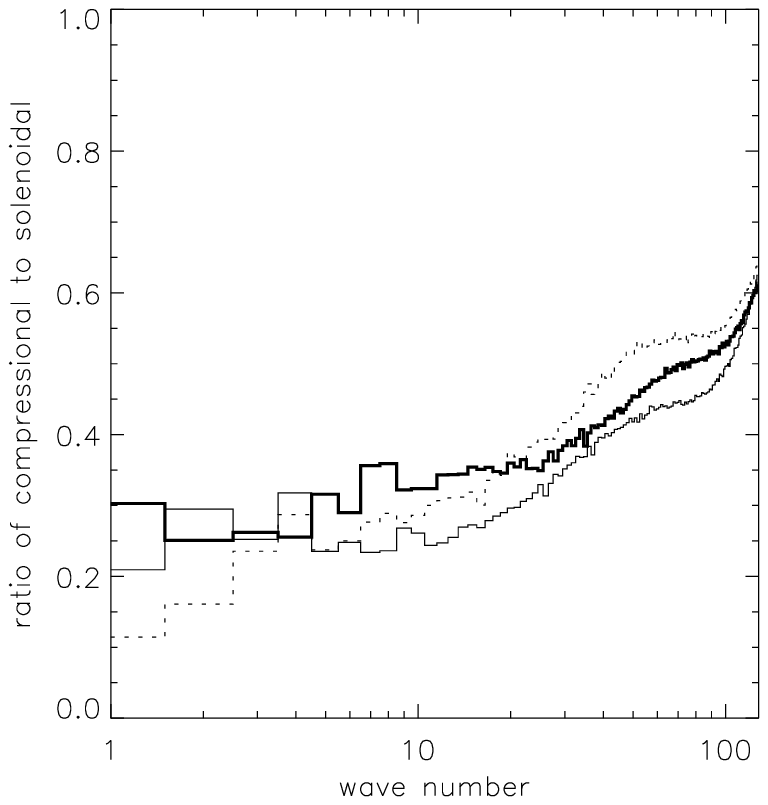}
\end{minipage}
\hfill
\begin{minipage}[b]{.49\linewidth}
\includegraphics[width=\linewidth]{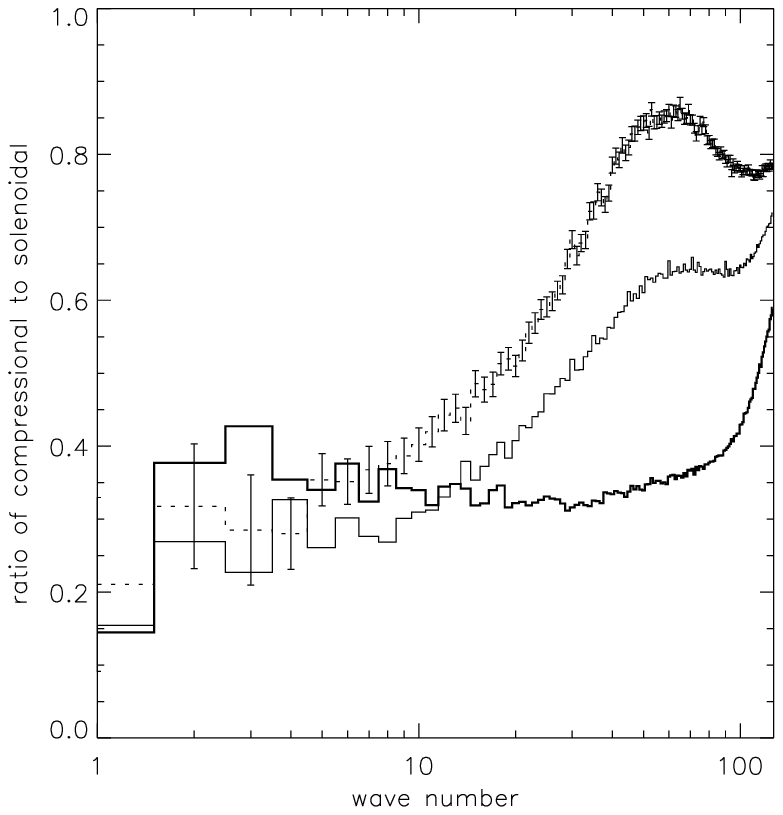}
\end{minipage}
\caption[Velocity power spectrum evolution in molecular turbulence]%
{Evolution of the velocity power spectrum (ratio of compressional to
solenoidal energy spectra) for \molt\ on the left and \isot\ on the
right.  Data at three instances are displayed: thick lines: $t=30$~yr;
thin lines: $t=300$~yr; dashed lines: $t=600$~yr.  See also
Fig.~\ref{eng:fig2}.  Typical amount of variability
due to the limited
number statistics can be assessed from the error bars plotted for the
$t=600$~yr IT case: values for large scales have smaller
statistical significance as there are fewer large scale modes (the
error bar for $k=1$ isn't shown but is large since it is represented 
by a single value).
}
\label{eng:fig2a}
\end{figure}

\begin{table}
\begin{center}
\begin{tabularx}{.5\linewidth}{lcc}
Time     & \molt    & \isot \\
\hline
$30$~yr  & $-2.91\pm0.06$  & $-2.88\pm0.06$ \\
$300$~yr & $-3.07\pm0.07$  & $-3.07\pm0.07$ \\
$600$~yr & $-3.16\pm0.07$  & $-3.16\pm0.07$ \\
\end{tabularx}
\end{center}
\caption[Linear fit coefficients $n$ of total energy power spectra]
{Coefficients, $n$, of linear fits to the total energy power spectra ($\log P \propto n \log k$) 
for the two cases of molecular gas (MT) and gas with an isothermal equation of state (IT). The indices 
were measured over the interval $k\in[8,64]$ and demonstrate a systematic steepening of 
the curves with time.}
\label{eng:tab1}
\end{table}

The main result is that there is no significant difference between the evolutions of the total 
power spectra of \molt\ and \isot. Since the turbulence is not driven, the statistical
properties of the fluid continue to change (as opposed to the dynamical
equilibrium reached in driven turbulence simulations). 

\subsection{Compressional and Solenoidal Power Spectra}
\label{sec:com}

The power in compressional and solenoidal modes begins in a 1:2 ratio, as noted above.
Intuitively, it might seem that the power ratio should scale with Mach number since a flow 
with zero Mach number is incompressible (all kinetic energy is in vortices) while a flow 
with a very large Mach number should be highly compressible \citep{Elmegreen04}. In our
case, however, 
solenoidal and compressional modes are coupled and can exchange energy in both 
directions. Of special relevance here is how the evolution of vorticity,
$\bm{w}\equiv[\nabla\times\bm{v}]$, is coupled with the divergence, 
$(\nabla \cdot \bm{v})$,
\begin{equation}
\label{eng:eq6}
\begin{split}
  \frac{\partial \bm{w}}{\partial t} + \left(\bm{v} . \nabla\right) \bm{w} =& 
  \left(\bm{w} . \nabla\right)\bm{v} - \bm{w} \left(\nabla . \bm{v}\right)\\
  -& \left[\nabla\left(\frac{1}{\rho}\right)\times\nabla P\right] 
  + \nu \nabla^2 \bm{w},
\end{split}
\end{equation}
where $\rho$ is the mass-density of the fluid, and $P$ is pressure.  The 
first term on the right hand side of equation~(\ref{eng:eq6}) describes the vorticity
generation due to tilting and stretching. The second term couples generation
of the vorticity and divergence (if there is a negative divergence in the
fluid, it will tend to shrink a fluid parcel, decreasing its moment of
inertia, and, as a result, angular momentum conservation will give rise to 
vorticity), and the third term describes baroclinic vorticity generation.  If
turbulence is modelled as barotropic or isothermal, baroclinic
vorticity generation is suppressed \citep{Elmegreen04}.

As displayed in Figs.~\ref{eng:fig2a}, the ratio of compressional to 
solenoidal energy does not remain constant across the range of scales, with large scales 
being more solenoidal and small scales slightly more compressive. This is not a surprising 
result as compressive motions are associated with shocks which are only a few zones wide 
(as allowed by artificial viscosity \cite[see, e.g.,][]{Stone92I}) and thin shells from 
subsequent radiative cooling. The shells  break up due to various instabilities 
\citep{Vishniac83,M-MML93,Vishniac94,Dgani96}.

It is remarkable that small scale compressional motions develop much faster in the \molt\ case 
than in the \isot\ case (we thank M.~Bate for pointing this out).  This effect can be partly due 
to additional thermal instabilities \citep{Smith03} that contribute to the rapid braking of 
shocks in the \molt\ case, but is more likely to be associated with the higher sound 
speed in the dissociated gas that allows compressive waves to steepen faster.  
In the \isot\ case, large scale shocks survive longer and compressional energy gradually 
shifts from large to  small dissipative scales.  Another possibility is additional
baroclinic vorticity generation in the \molt\ case, as follows from the analysis of the 
equation for vorticity evolution, Eq.~(\ref{eng:eq6}). Statistical results for shocks and 
vorticity will be discussed in detail in \S~\ref{ssec:vels}.

Once developed, the IT becomes more compressive at high wavenumbers than MT, remaining, 
however, predominantly solenoidal in both cases. 
The inequality $P_c/P_s < 1$ has also been found in a simulation of driven supersonic 
magnetised turbulence \citep{Boldyrev02b,Padoan02}. Given that the total (solenoidal plus
compressional) energy is the same in both simulations (see Fig.~1), we attribute this 
difference to baroclinic vorticity generation in MT.

A related result was also found in simulations of driven turbulence with a
multiple power-law cooling function \citep{Vazquez96}. The initial turbulence was
purely solenoidal but was driven by purely compressible forcing.  Vorticity production
in this case was found to depend heavily on the additional terms (including Coriolis 
force on large scales, and Lorentz force on small scales).

This interplay between modes was also noted by \cite{Balsara04}.
There, even though the medium is driven by highly compressible motions, the 
kinetic energy  is mainly concentrated in solenoidal rather than the 
compressible motions. This behaviour was shown to arise from the interaction 
of strong shocks with each other and with the interstellar turbulence they 
self-consistently generate.

\section{SIMULATED STRUCTURE} 
\label{sec:vsh}

\subsection{Density PDFs}
\label{ssec:dens}

The basic statistical property of a molecular cloud model or, indeed, observational data 
is the distribution of density. While the observational picture is essentially 
two dimensional, our simulations provide detailed information about the density 
distribution in 3D.  This distribution can be described in several ways. 
The fractional volume distribution is
\begin{equation}
\label{dens:eq1}
   \text{d}{\cal P}_V = \Omega_V\text{d}\rho,\qquad\Omega_V = \frac{1}{V_0}
   \frac{\text{d}V}{\text{d}\rho},
\end{equation}
where $\text{d}{\cal P}_V$ is the probability to find a value of density
$\rho\in[\rho,\rho+\text{d}\rho]$ and $V_0$ is the total volume. Similarly,
the fractional mass distribution is
\begin{equation}
\label{dens:eq2}
   \text{d}{\cal P}_M = \Omega_M\text{d}\rho,\qquad\Omega_M = \frac{1}{M_0}
   \frac{\text{d}M}{\text{d}\rho},
\end{equation}
where $M_0=\langle\rho\rangle V_0$ is the total mass and  $\langle\rho\rangle$ is 
the volume averaged density.

To describe molecular turbulence, we introduce the fractional {\it molecular} mass,
\begin{equation}
\label{dens:eq3}
   \text{d}{\cal P}_\frak{M} = \Omega_\frak{M}\text{d}\rho,\qquad
   \Omega_\frak{M} = \frac{1}{\frak{M}_0}
   \frac{\text{d}\frak{M}}{\text{d}\rho},
\end{equation}
where $\frak{M}=(2/1.4)f\rho\text{d}V$ is the mass of \ch{H_2} 
(here, $fn\times2m(\ch{H})$ is the mass of \ch{H_2}, and $\rho = 1.4nm(\ch{H})$ is
the density of the mixture: \ch{H}, \ch{H_2} and 10\% of \ch{He}) and $\frak{M}_0$
is the total mass of \ch{H_2}.

The distributions given by
equations~(\ref{dens:eq1})--(\ref{dens:eq3}) are normalised:
\begin{equation}
\label{dens:eq4}
   \int_{-\infty}^{\infty}\Omega_\Sigma = 1,\qquad \Sigma=\{V,M,\frak{M}\}
\end{equation}
where $\Sigma$ represents any
index. Note that the
$\Omega_\Sigma$ are not independent characteristics of the density 
distribution since it follows from equations~(\ref{dens:eq1}) and 
(\ref{dens:eq2}) that
\begin{equation}
\label{dens:eq5}
   \Omega_M(\rho) = \frac{\rho}{\langle\rho\rangle}\Omega_V(\rho).
\end{equation}

Previous analyses of $\Omega_V$ and $\Omega_M$ in simulations of 
compressible hydrodynamic supersonic turbulence have 
demonstrated that when the equation of state is approximately isothermal, the density
distributions are close to log-normal, i.e. that the logarithm of
density has a probability distribution function (PDF) that is a Gaussian
\citep{Vazquez94, Padoan97, Passot98, Scalo98, Ostriker01}.  
\citet{Scalo98} found that a log-normal density PDF should only
occur at (1) low Mach numbers or (2) for any value of the Mach number when
the polytropic index $\gamma$ is equal to unity (i.e. in the isothermal case).
Otherwise, they found density PDFs to be well approximated by power-laws over
wide regions of $\log(\rho)$ variation.  The simulations performed 
by \citet{Scalo98} were of two-dimensional, driven, galactic turbulence, 
and included a variety of physical processes expected on such 
scales, including self-gravity, magnetic fields, rotation, heating and 
cooling.  

It is not clear, however, whether the polytropic 
index (ratio of specific heats) alone is responsible for the shape 
of density PDFs.  \citet{Li03} studied a number of cases of turbulence
with an adiabatic equation of state with different polytropic indices.
In the case of driven hydrodynamical turbulence, it was found that
both mass and volume PDFs show imperfect log-normal distributions.
During dynamical evolution, the PDFs in the non-isothermal cases were
found to develop only small
deviations from Gaussian fits, while PDFs
in the isothermal case ($\gamma=1$) remain very well fitted by a
Gaussian.
Simulations of three-dimensional, supernova-driven, galactic
turbulence by \citet{M-MML05} also show generally Gaussian results
even though heating and cooling are included explicitly,
consistent with the results of \citet{Klessen00p} who studied the PDFs of isothermal, self-gravitating
turbulence. However, he found that
the moments of the 
density PDFs vary as collapse proceeds. Hence, self-gravity is also an
important factor in shaping density PDFs. It is
plausible that other physical processes can affect the shape of the 
density PDF.

For parameterisation of PDFs with log-normal distributions, it is useful to define a 
dimensionless variable $\chi\,=\,\ln (\rho/\avrg{\rho})$. Substituting into
\begin{equation}
   \frac{\text{d}{\cal P}_{\Sigma}}{\text{d}\rho} 
     = \Omega_{\Sigma}(\rho)
\end{equation}
yields
\begin{equation}
\label{dens:eq8}
   {\cal P}_{\Sigma}(\chi) \frac{1}{\rho} = \Omega_{\Sigma}(\rho).
\end{equation}
If $\Omega_{\Sigma}(\rho)$ is a log-normal distribution, 
then ${\cal P}_{\Sigma}(\chi)$ should be the normal (Gaussian) distribution
\begin{equation}
\label{dens:eq9}
   {\cal P}_{\Sigma}(\chi) = \frac{1}{\sqrt{2\pi}\sigma_{\Sigma}}
   \exp \left[ - \frac{(\chi - \chi_{\Sigma})^2}{2\sigma^2_{\Sigma}} \right].
\end{equation}
Following the work of \citet{Li03}, we relate ${\cal P}_M(\chi)$ 
and ${\cal P}_V(\chi)$, using equations~(\ref{dens:eq8}) and (\ref{dens:eq5}):
\begin{equation}
\label{dens:eq10}
   {\cal P}_M(\chi) = e^{\chi} {\cal P}_V(\chi).
\end{equation}
For the functional dependence given by equation~(\ref{dens:eq9}) to be valid
for both PDFs, parameters of the normal distributions have to satisfy
the following relations due to Eq.~(\ref{dens:eq10}):
\begin{gather}
\label{dens:eq11}
   \sigma_V = \sigma_M \equiv \sigma,\\
\label{dens:eq12}
   \chi_M = \chi_V + \sigma^2,\\
\label{dens:eq13}
   \chi_V = -\frac{\sigma^2}{2}, \quad
   \xrightarrow{\text{(\ref{dens:eq12})}} \quad
   \chi_M = \frac{\sigma^2}{2},
\end{gather}
as noted by \citet{Ostriker01}.

In terms of the PDFs, the distributions  $\Omega_{\frak M}(\rho)$ and $\Omega_V(\rho)$ are
related by
\begin{equation}
\label{dens:eq14}
   \Omega_{\frak M} = \frac{f\rho}{\avrg{f\rho}}\Omega_V,
\end{equation}
where $f$ is molecular fraction ($fn$ is abundance of \ch{H_2}), 
$\avrg{f\rho}=(1.4/2){\frak M}/V$.
Using equations~(\ref{dens:eq14}), (\ref{dens:eq5}) and (\ref{dens:eq8}) we
can relate the normal distributions for mass and molecular mass through
\begin{equation}
\label{dens:eq15}
   {\cal P}_{\frak M}(\chi) = \frac{f\avrg{\rho}}{\avrg{f\rho}}{\cal P}_M(\chi).
\end{equation}

\subsection{The Effective Specific Heat Ratio}
\label{ssec:gamma}

\begin{figure}
\begin{minipage}[b]{.48\linewidth}
\centering\includegraphics[width=.9\linewidth]{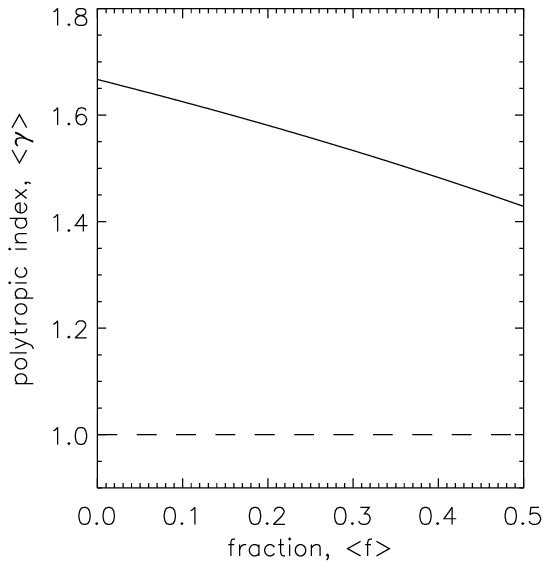}
\caption[Average polytropic index as a function of $\avrg{f}$]%
{The average specific heat ratio  $\avrg{\gamma}$ employed by the code
as a function of the average molecular fraction 
$\avrg{f}$ in a warm molecular/atomic gas mixture (solid line);
the isothermal gas case $\gamma=1$ is also shown (dashed line).}
\label{dens:fig1}
\end{minipage}
\hfill
\begin{minipage}[b]{.48\linewidth}
\centering\includegraphics[width=\linewidth]{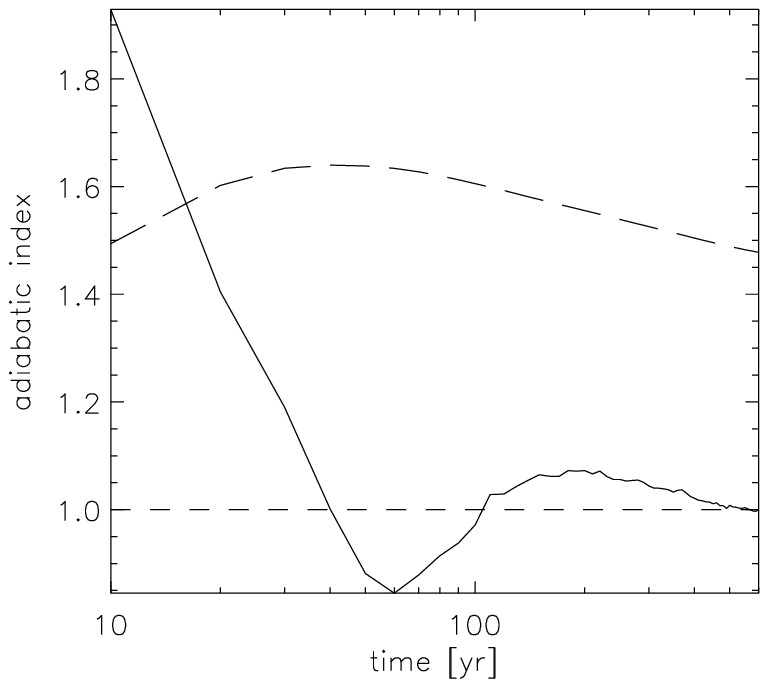}
\caption[Change of the average polytropic index with time]%
{Long dashed line: $\avrg{\gamma}$ the average specific heat ratio 
as a function of time in the \molt\ simulation; Continuous line: $\gamma_p$ the effective specific 
heat ratio as a function of time; short dashed line: the isothermal value (1.0)}
\label{dens:fig2}
\end{minipage}
\end{figure}

The actual ratio of specific heats, $\gamma$, as calculated for each zone
in our \molt\ simulation, is a monotonic function of the molecular fraction, $f$,
\begin{equation}
\label{dens:eq7}
   \gamma=(5.5-3f)/(3.3-f),
\end{equation}
(including 10\% helium). As shown in Fig.~\ref{dens:fig1}, $\gamma$ may vary between 5/3 and 10/7. 

\begin{figure}
\centering
\begin{minipage}[b]{\linewidth}
\centering\includegraphics[width=.6\linewidth]{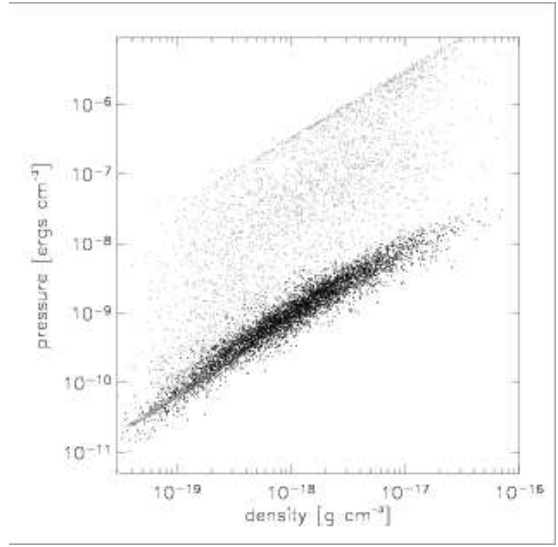}
\caption[Scatter plots: pressure v density]%
{Scatter plots: logarithm of pressure vs logarithm of density.  
Grey points correspond to the time $t=10$~yr, and black points
$t=600$~yr; linear fitting $\log(P) \propto \log (\rho)$ has a coefficient close to 1.0 at 
later times, see Fig.~\ref{dens:fig2} }
\label{dens:fig2a}
\end{minipage}
\end{figure}

We can define a spatially-averaged specific heat ratio by 
substituting $f \longrightarrow \avrg{f}$ in Eq.~(\ref{dens:eq7}).
With these definitions, $\avrg{\gamma}$ does not depend on the gas excitation. 
Hence, strictly speaking, it can serve as a good approximation only for a gas 
at a high  temperature, to ensure that the rotational degrees of freedom are 
fully excited and the classical approximation for specific heats, which we employed 
in our model, remains valid.  For molecular hydrogen, this is true
for temperatures $T>85$\,K.  As shown in Fig.~\ref{dens:fig2},
$\avrg{\gamma}$ varies considerably during the simulation.

To compare to the isothermal simulation, we define a polytropic index $\gamma_p$.
This is the {\em effective} specific heat ratio including radiative cooling. 
We define it as the linear fit coefficient for the dependency
\begin{equation}
\log(P) = \gamma_p \log(\rho) + \text{const},
\end{equation}
where $P$ is pressure and $\rho$ is total mass density.
In Fig.\ref{dens:fig2a}, we display the $P-\rho$ data at two times.
At early times (grey cells), the dependency displays a great deal of scatter
but is effectively bound between two $\gamma_p=1$ lines which, overall, results in
the linear fitted value being close to 2, see Fig.~\ref{dens:fig2}.  The upper line 
$\gamma_p=1$ is formed by the highest temperature regions; this temperature corresponds to $\sim$\,8000K,
the post-shock temperature following rapid atomic cooling but before molecular cooling has become
effective. The lower line is a reminder of the original $T=100$~K isothermal 
conditions.  At later times (black cells in Fig.~\ref{dens:fig2}) the scatter is reduced, 
and it indicates an effective isothermality of the gas.

\subsection{PDFs: analysis} 
\label{ssec:pdf}

In fully molecular gas $f(\rho) = \avrg{f} = 0.5$ and ${\cal P}_{\frak M} = {\cal P}_M$.
Hence, differences between these two distributions are only going to be noticeable when strong 
shocks dissociate a considerable amount of \ch{H_2}. As shown in Fig.~\ref{dens:fig2}, this 
occurs in the first 100~yr of the simulation.  Over the next 500~yr, owing to the  reformation 
of molecules, we can expect large deviations of ${\cal P}_{\frak M}$ from  ${\cal P}_M$ (see also 
Fig.~4 of Paper~I).  Finally, when dissociative shocks disappear and the molecular fraction 
distribution flattens, ${\cal P}_M$ and ${\cal P}_{\frak M}$ should converge.

\begin{figure}
\begin{minipage}[b]{.48\linewidth}
\includegraphics[width=\linewidth]{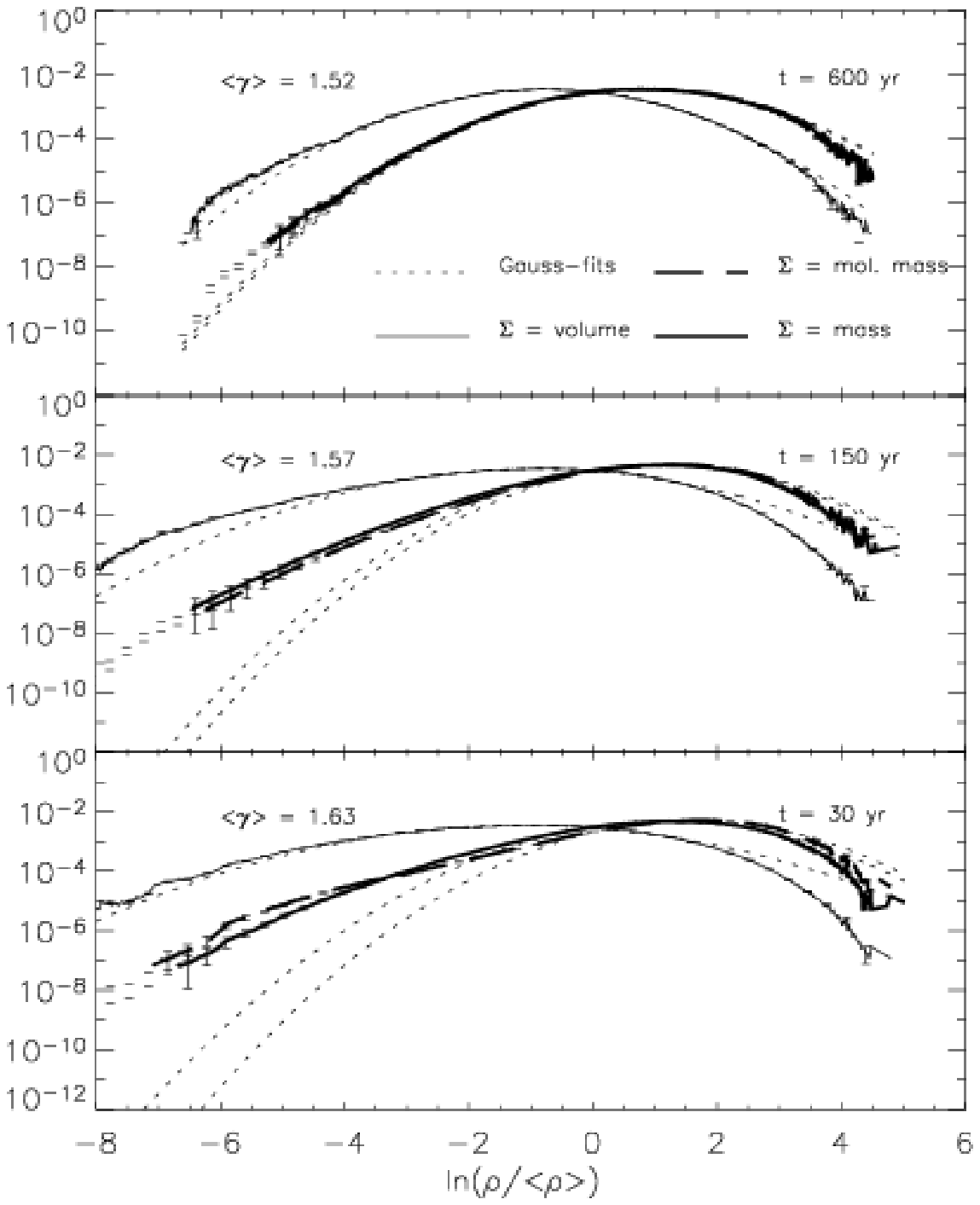}
\end{minipage}
\hfill
\begin{minipage}[b]{.48\linewidth}
\includegraphics[width=\linewidth]{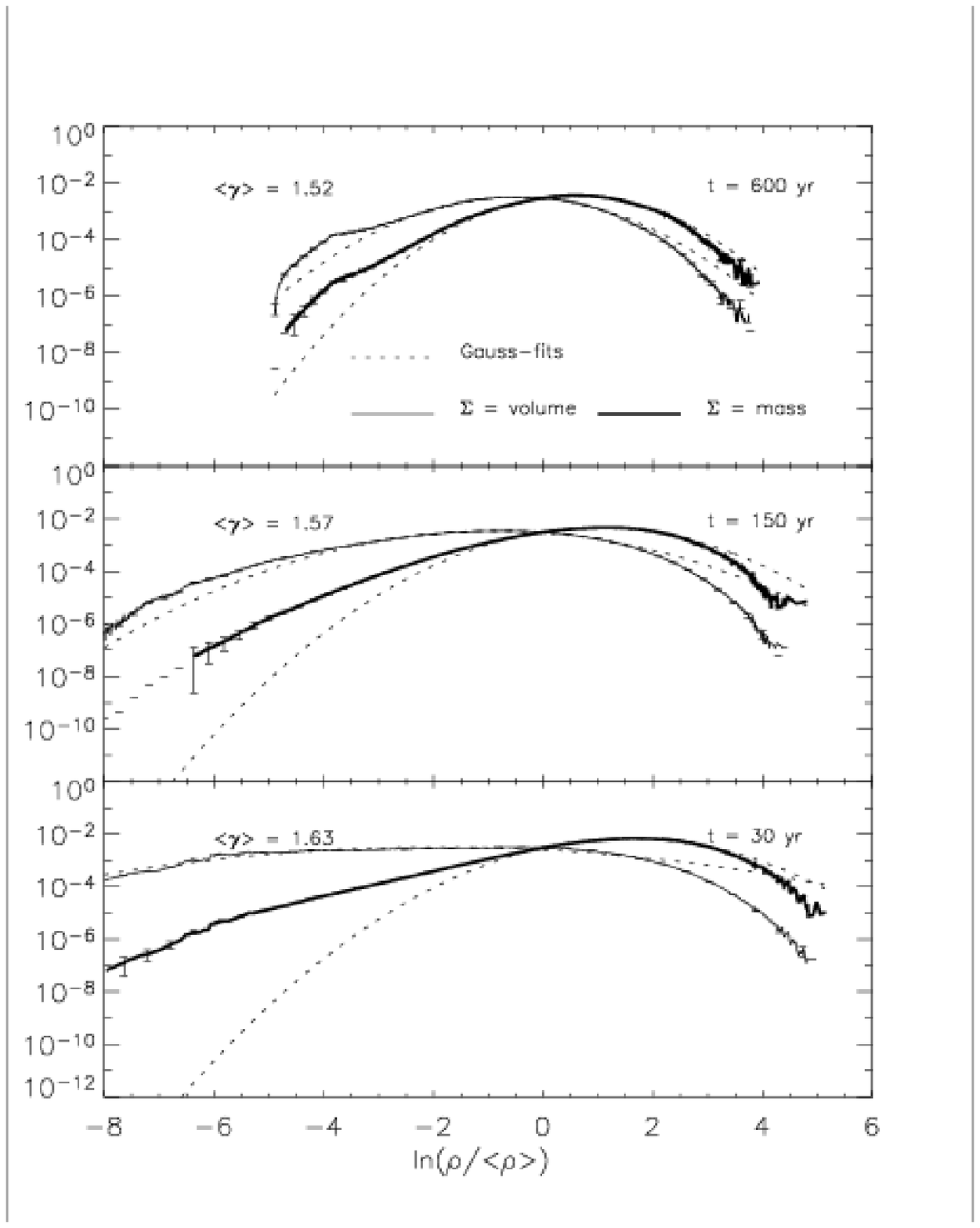}
\end{minipage}
\caption[Evolution of PDFs of the density distribution in
turbulence]%
{Evolution of PDFs of the density distribution during the simulation of decaying, 
high-speed, molecular turbulence including heating and cooling (\molt, left panel) 
and isothermal turbulence (\isot, right panel).  Note power-law wings of the mass-weighted
PDFs, and more consistent log-normal shape of the volume-weighted PDFs.
To demonstrate statistical significance of the
data points we plot error bars for every 20th point of each PDF curve.  When statistical significance 
of a PDF point is small (i.e., it is derived from a single data value) the curve is traced with a
small horizontal dashes (see extreme of the wings of the mass-weighted PDFs). 
}
\label{dens:fig3}
\end{figure}

The volume-weighted 
density PDFs in the \molt\ run are found to be close to log-normal, as shown on the 
series of plots in Fig.~\ref{dens:fig3} (left panel). 
Volume-weighted PDFs of density for
Burgers' turbulence (completely shock dominated turbulence)
have been analytically shown to 
have power-law asymptotics \citep{Frisch01,Frisch01a}.
It is clear from the error bars on  
the plots that we can't make a definitive statement about the asymptotic behaviour due to the lack of
numerical resolution,
but our results do not show clear evidence for this behavior.  
Numerical observations of PDF asymptotics require 
extremely high resolution even in one-dimensional cases \citep[see,
e.g.,][]{Gotoh93,Gotoh98}.  
We note however, that, during the early hypersonic evolution of both the
\isot\ and \molt\ cases, the {\em mass}-weighted PDFs display clear
power-law tails at low densities.  We remain uncertain how to
interpret this result.

To determine how much the deviation from a 
log-normal distribution is influenced by the high and variable value of the specific heat ratio and the 
introduction of explicit heating and cooling, we contrast these results with PDFs 
from \isot\ (right panel of Fig.~\ref{dens:fig3}), which has exactly the
same initial physical conditions and numerical properties.
The PDFs correspond to the same moments of time.
Both data sets suggest that deviations from log-normal distributions
are more significant at earlier times.  As turbulence decays, the
distributions become more consistent with log-normal properties (see
equations~(\ref{dens:eq11}) --- (\ref{dens:eq13}) ). 
We find that the PDF 
of density from the \molt\
simulation is 
actually more consistent with a log-normal distribution than
that from \isot.

\begin{figure}
\begin{minipage}[b]{.48\linewidth}
\includegraphics[width=\linewidth]{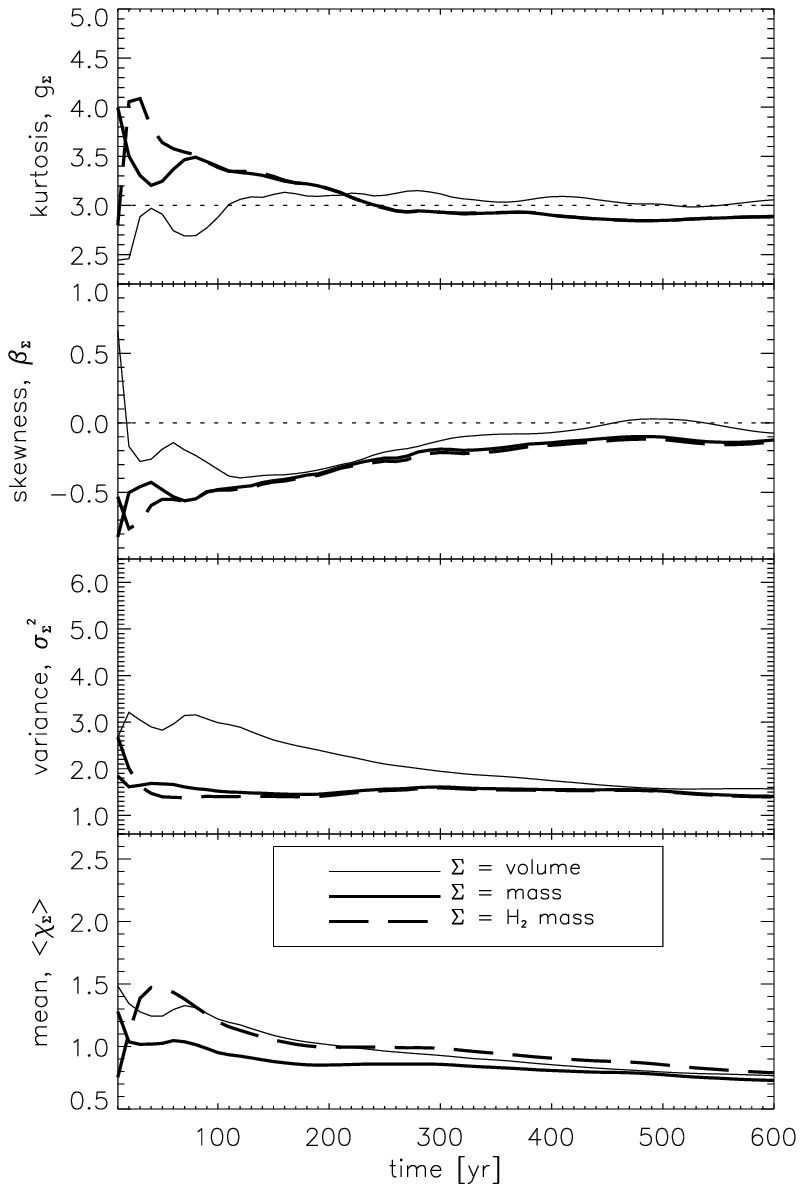}
\end{minipage}
\hfill
\begin{minipage}[b]{.48\linewidth}
\includegraphics[width=\linewidth]{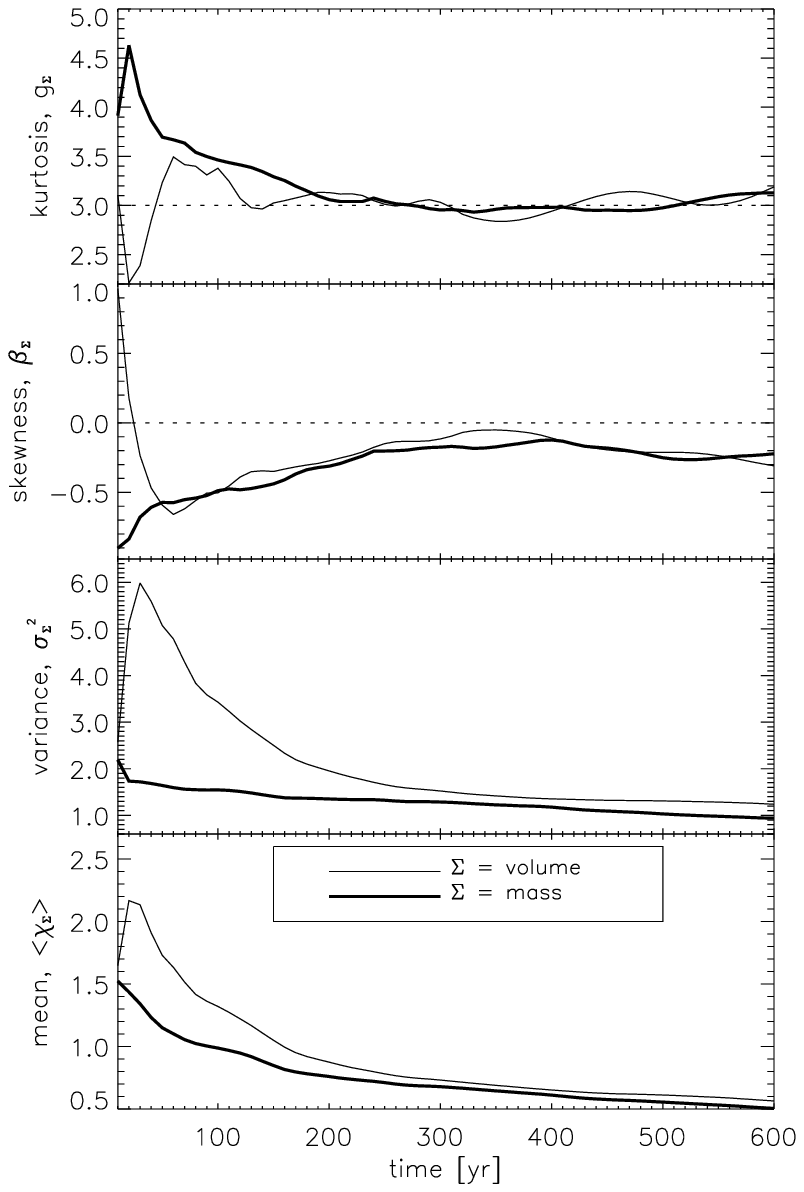}
\end{minipage}
\caption[Moments of the density distributions as a function of time]%
{First moment (mean) $\chi_\Sigma$, second moment (variance) $\sigma^2_\Sigma$, third moment (skewness)
$\beta_{\Sigma}$, and fourth moment (kurtosis) $g_{\Sigma}$ of 
${\cal P}_\Sigma$ as a function of time in the \molt\ (left) and \isot\ (right) simulations.
The volume mean, $\chi_V$ is plotted with inverse sign.  A Gaussian distribution has skewness, $\beta=0.0$,
and kurtosis, $g=3$, as indicated by the dotted lines.}
\label{dens:fig4}
\end{figure}

The evolution of the first four moments (mean $\chi_\Sigma$, variance
$\sigma^2_\Sigma$, skewness, $\beta_{\Sigma}$, kurtosis, $g_{\Sigma}$) of the 
PDFs are displayed in Fig.~\ref{dens:fig4}. 
The variance, $\sigma^2_V$, exhibits a large deviation from $\sigma^2_M$ early 
on during both simulations.  However, in the \molt\ case the 
deviation is less significant.  In both cases, $\sigma^2_V$ rises sharply 
when shocks form and distort the initially homogeneous density field.
At the beginning of the simulation, in molecular gas the density contrast 
is less than in isothermal gas because $\gamma>1$. 
As shown in Fig.\ref{dens:fig2}, a significant hot gas phase
is produced.
This can qualitatively 
explain smaller values of $\sigma^2_V$ in a molecular turbulent flow.
The variances  
$\sigma^2_M$ and $\sigma^2_{\frak M}$ are approximately constant during the
simulations.  In the \molt\ simulation, parameters of mass and
molecular mass distributions converge, as expected from the
above analysis.

To measure deviation of the log-distribution from gaussianity we also compute higher order moments --- 
skewness, $\beta$, and kurtosis, $g$,
\begin{equation}
\beta = \frac{\mu_3}{\mu_2^{3/2}},\quad
g = \frac{\mu_4}{\mu_2^2},
\end{equation}
where
\begin{equation}
\mu_n = \avrg{(\chi - \avrg{\chi})^n}
\end{equation}
is a central moment of $n$-th order.  For a normal distribution, skewness, a degree of asymmetry of the 
distribution, $\beta = 0$, and kurtosis, a degree of peakedness of the distribution, $g=3$.  For
an exponential PDF the corresponding values are $\beta=2$, $g=6$.
As seen from plots on Fig.~\ref{dens:fig3} a significant deviation from Gaussian shape 
occurs only at earlier times. However, skewness remains negative throughout, indicating longer
`tail' on the side of low densities.  It can be attributed to the fact that highest values of density
are limited by the highest Mach number, whereas the lowest density does not depend on properties of 
individual shocks directly.

These data suggest that $\avrg{\gamma}$ only slightly influences
the PDFs of the density distributions. 
Although these results are more generally applicable than specifically
to large-scale, cold molecular clouds, where we would expect
values of polytropic index $\gamma\approx 1$, it is 
remarkable that an entirely different equation of state 
leads to
qualitatively and quantitatively very similar statistical properties.

\subsection{Velocity: vorticity and shock fields} 
\label{ssec:vels}

As remarked above, the assumption of isothermal or isentropic flow stops baroclinic 
vorticity creation.  It also forces helicity, $\omega = \avrg{(\bm{v} \cdot \bm{w})}$, 
to be conserved in an unphysical way in general compressible flows. In this sub-section, 
we analyse in detail the differences in the amount of vorticity and shocks  in the 
\molt\ and \isot\ cases.

\begin{figure}
\begin{minipage}[b]{.5\textwidth}
\centering\includegraphics[width=0.95\textwidth]{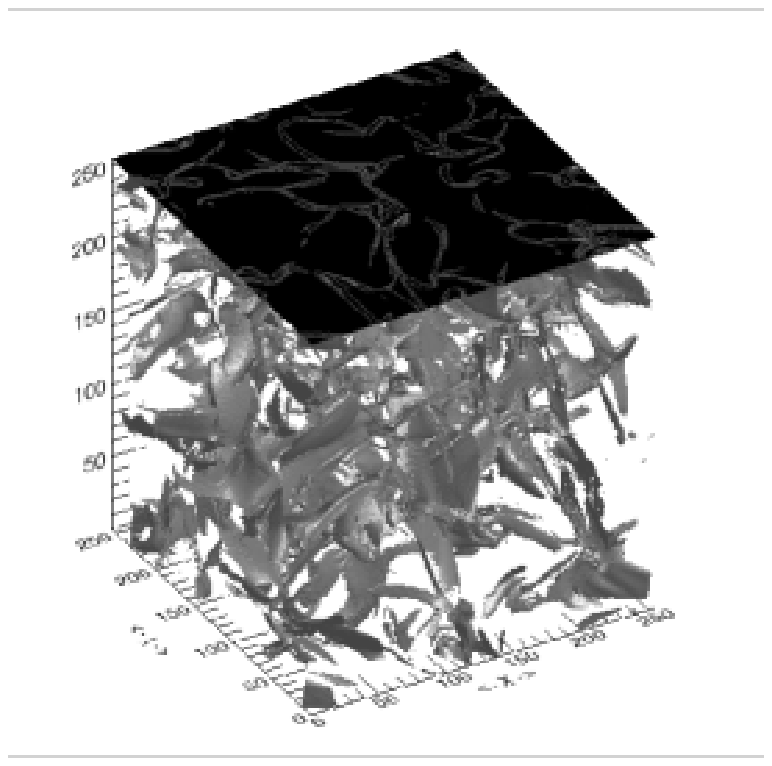}
\end{minipage}\hfill
\begin{minipage}[b]{.5\textwidth}
\centering\includegraphics[width=0.95\textwidth]{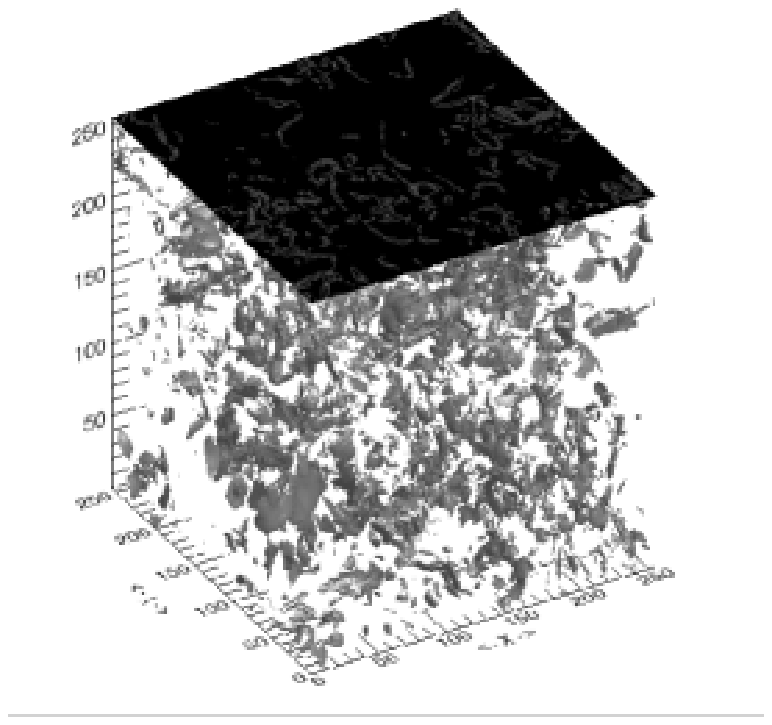}
\end{minipage}
\caption[3D: vorticity and shocks at $t=10$~yr and $t=300$~yr]%
{{\bf Left cube:} Surfaces of isovorticity, $\big|[\nabla\times\bm{v}\big|=\text{const}$ 
at 10~yr after the start of the \molt\ simulation. Shocks have just formed and large 
structures are present. The black upper face of the cube contains a contour plot of the 
top slice of vorticity.
{\bf Right cube:} surfaces of isovorticity at the later time
of 300~yr when the shocks have fragmented.}
\label{vels:fig1}
\end{figure}

The vorticity distribution is displayed in Fig.~\ref{vels:fig1}.
In isotropic, incompressible, turbulent flow, high vorticity exists in
thin coherent tubes, as found in direct 
numerical simulations (DNS) with high resolution
\citep[see, e.g.,][]{Lesieur03, Titon03}.  The vorticity tubes have a diameter 
of a few Kolmogorov dissipative scales and a length of the order of
the integral scale (i.e. the driving scale).
Centres of the vorticity tubes form
one dimensional filament-like structures 
that are the centres of energy dissipation of the flow. 

\begin{figure}
\begin{minipage}[b]{.5\textwidth}
\centering\includegraphics[width=0.95\textwidth]{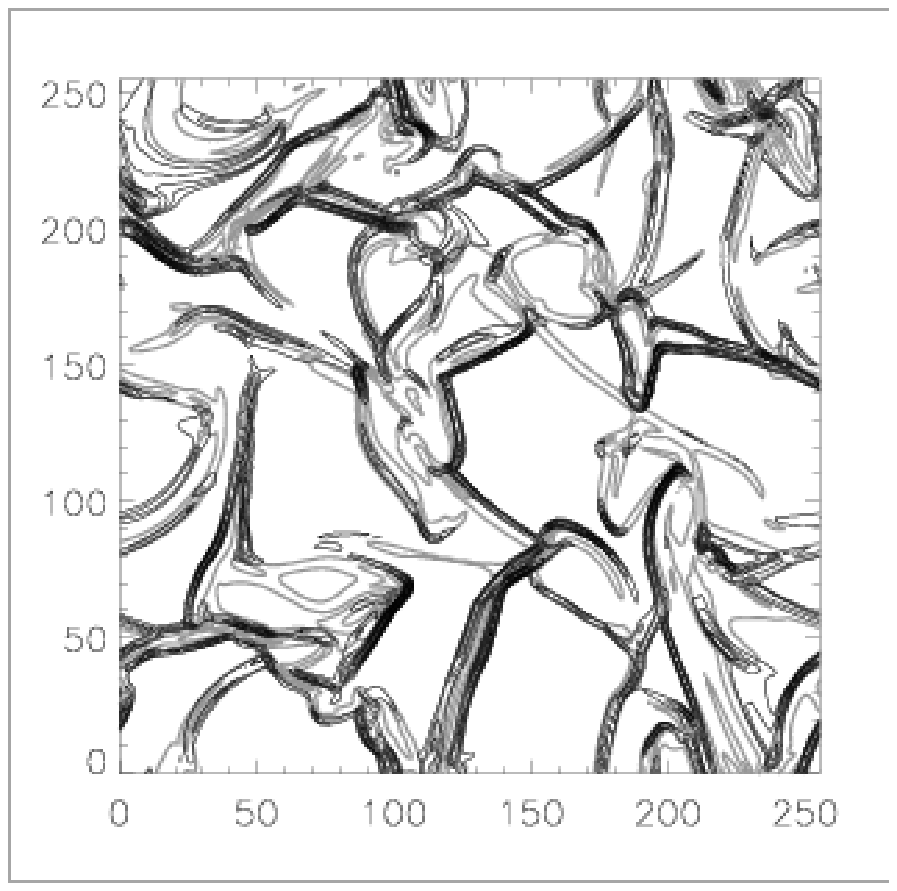}
\end{minipage}\hfill
\begin{minipage}[b]{.5\textwidth}
\centering\includegraphics[width=0.95\textwidth]{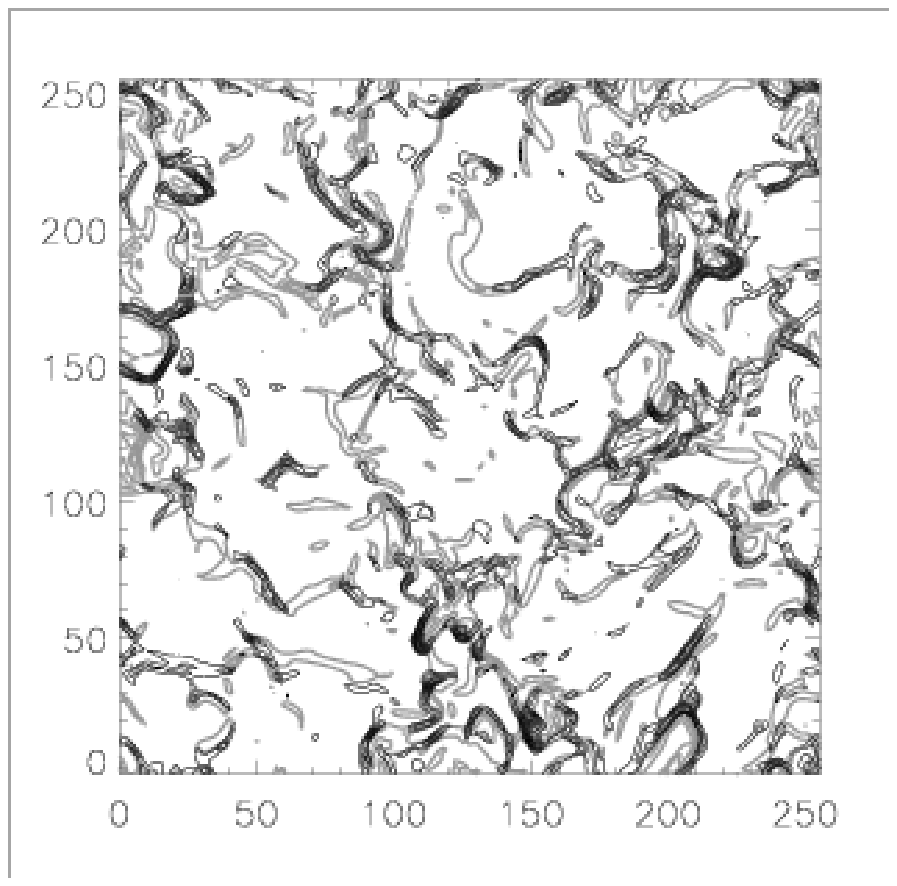}
\end{minipage}
\caption[Contours: vorticity and shocks at $t=10$~yr and $t=300$~yr]%
{Contour plots from the \molt\ case of negative velocity divergence (black)
and contours of the vorticity field (grey). Note that the vorticity is concentrated around 
the shocks. The plots display random slices from the data cubes at 
$t=10$~yr (left) and $t=300$~yr (right).}
\label{vels:fig2}
\end{figure}

In compressible flow, on the other hand, vorticity structures form sheets and spirals 
\citep{Porter02}.  In our simulations of decaying supersonic turbulence,
we find that the vortex structures (isosurfaces of high vorticity) form
``open'' surfaces.  The size of these surfaces is closely correlated with
the size of shock surfaces (Fig.~\ref{vels:fig2}).
Isosurfaces of strong vorticity become smaller with time,
with a characteristic size of order of the shock size, as displayed in
Figs.~\ref{vels:fig1} and \ref{vels:fig2}.

Statistical analysis of the vortex structures reveals that vorticity components as well 
as velocity gradients have PDFs of exponential type, i.e., proportional to 
$\exp\{-|X|\}$, in  incompressible turbulent flow. This is instead of the Gaussian
distributions proportional to $\exp\{-|X|^2\}$ that
would be expected for statistically independent structures 
due to the Central Limit Theorem. In laboratory experiments and DNS, 
regions of high vorticity correlate with regions of low 
pressure \citep[][p.196]{Lesieur97}.
The exponential form of the PDF of vorticity 
is a manifestation of the intermittency  of energy dissipation.  
If dissipative structures
were distributed homogeneously, the PDF would be given by a Dirac
delta function, or a narrow Gaussian distribution in a DNS.  
Therefore, a deviation of the PDF of dissipative structures from the
Gaussian shape is a direct indication of the intermittency of turbulence.

\begin{figure}
\begin{minipage}[b]{.5\textwidth}
\centering\includegraphics[width=0.99\textwidth]{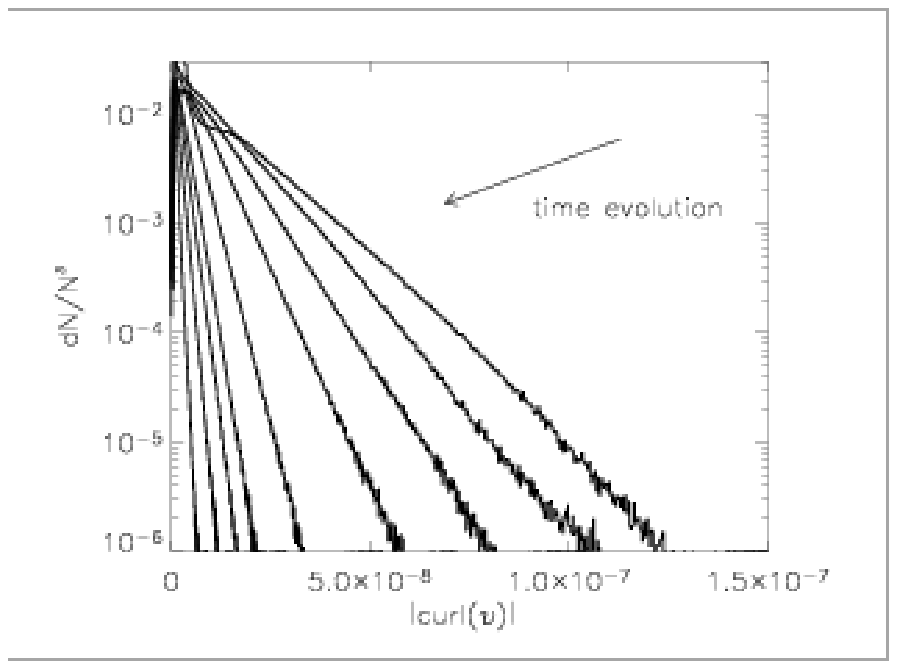}
\end{minipage}\hfill
\begin{minipage}[b]{.5\textwidth}
\centering\includegraphics[width=0.99\textwidth]{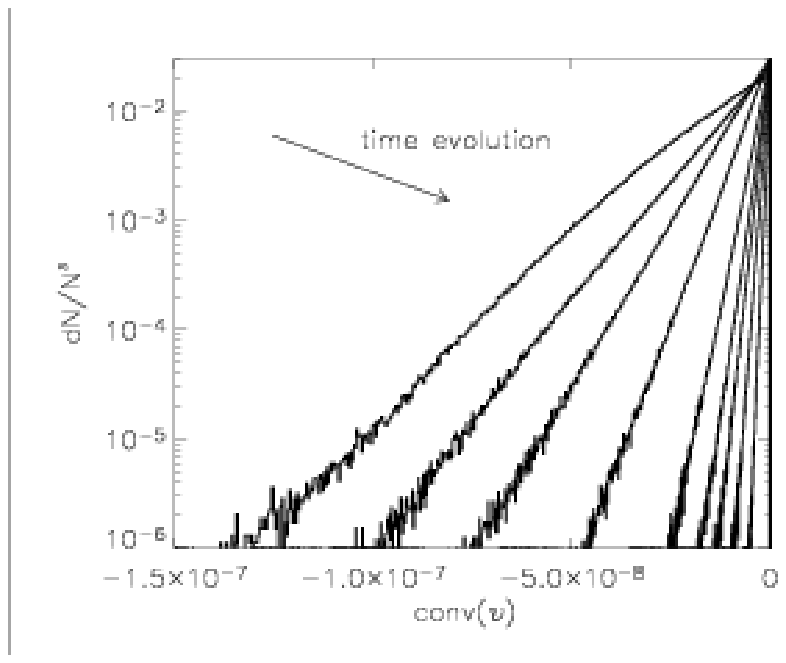}
\end{minipage}
\caption[Histograms of curl and div]%
{{\bf Left panel:} histograms of $\big|\text{curl}(v)\big|$ at different times (\molt\ simulation).
{\bf Right panel:} histograms of $\text{conv}(v)$ at different times (\molt\ simulation).  
The histograms correspond to the following simulation times:
10, 20, 30, 50, 100, 150, 200, 300, 600 [yr], with the direction of the time sequence 
marked by the arrows on the figures i.e. the distributions steepen.}
\label{vels:fig3}
\end{figure}

To study PDFs of shock waves we define the function of convergence, $\text{conv}(v)$,
\begin{equation}
\label{vels:eq1}
   \text{conv}(v) =
   \begin{cases}
   \left(\nabla \cdot \bm{v}\right), &\text{when}
   \left(\nabla \cdot \bm{v}\right)<0,\\ 
   0, &\text{otherwise}.
   \end{cases}
\end{equation}
A simple statistical analysis of the converging regions and vorticity reveals the expected 
exponential PDFs for both fields (Fig.~\ref{vels:fig3}).  Exponential laws for PDFs related to
convergence were reported before by \citet{Smith00a}, where the distribution of shock jump 
velocities in a single direction were found to be approximated by an exponential:
\begin{equation}
\label{vels:eq2}
   \begin{split}
   \text{d} N &\equiv\bigg[\text{number of zones with }
   \Delta v_i\in[\Delta v_i, \Delta v_i + \epsilon]
   \bigg]\\
   &\propto \exp\{ -C\Delta v_i \},
\end{split}
\end{equation}
where $C$ is a positive constant, and $\epsilon$ is a small discretisation parameter.

We find that the PDF of $\text{conv}(v)$, which is essentially a 3D generalisation of   
Smith et al.'s one dimensional jump velocity,  has the same exponential dependence,
\begin{equation}
\label{equ:sh3}
   \frac{\text{d} N}{N^3} \propto \exp\{ -\beta|\text{conv}(v)| \},
\end{equation}
where $\text{d} N$ is the number of zones with $\text{conv}(v)\in[\text{conv}(v),
\text{conv}(v)+\epsilon]$ and $N^3$ is the total number of zones (see 
Fig.~\ref{vels:fig3}).  We have chosen the discretisation parameter 
$\epsilon$ to create $500$ bins for our histograms.  

Vorticity PDFs are studied using the modulus of curl, 
$\big|\text{curl}(v)\big|\equiv\big|[\nabla\times\bm{v}]\big|$.  We find
that the PDF of curl behaves very similarly to the 
PDF of convergence but with different parameters:
\begin{equation}
\label{equ:sh4}
\frac{\text{d} N}{N^3} \propto \exp\{ -\alpha|\text{curl}(v)| \}.
\end{equation}

The coefficients $\alpha$ and $\beta$ change as the turbulence decays.  Both PDFs 
become progressively steeper with time, indicating that strong shocks and vorticity 
disappear. However, we find that the PDFs of convergence and vorticity do not change 
with time if we compensate for the average convergence and curl decay. The coefficients
$\alpha$ and $\beta$ of the distributions,
\begin{gather}
\label{vels:eq5}
   \frac{\text{d} N}{N^3} \propto \exp\{ \alpha\frac{|\text{curl}(v)|}
   {\avrg{|\text{curl}(v)|}} \}, \\
\label{vels:eq6}
   \frac{\text{d} N}{N^3} \propto \exp\{ \beta\frac{\text{conv}(v)}
   {\avrg{\text{conv}(v)}} \},
\end{gather}
do not significantly change during the simulation. The values of the coefficients for 
\molt\ and \isot\ simulations are given in Table~\ref{vels:tab1}.  These values may be significant
constants characterising decaying turbulence.

\begin{table}
\begin{center}
\begin{tabularx}{0.43\linewidth}{ccc}
            & \molt\           & \isot\         \\
\hline
$\alpha$    & $2.63\pm0.05$   & $2.61\pm0.06$ \\
$\beta$     & $2.48\pm0.05$   & $2.30\pm0.05$ \\
\end{tabularx}
\caption[Coefficients of compensated PDFs of convergence and curl]%
{Coefficients of compensated PDFs of convergence ($\beta$) and module of curl ($\alpha$). 
The error estimates are derived from the statistical variance of the distributions of the 
coefficients derived from different moments in time).}
\label{vels:tab1}
\end{center}
\end{table}

The data in Table~\ref{vels:tab1} suggest that the compensated PDF of vorticity is not
influenced strongly by a different equation of state, whereas the compensated PDF of convergence 
is steeper in the \molt\ case. A shallower PDF of convergence implies the existence of larger
quantities of strong shocks in the \isot\ case, which is not surprising. Shock jump conditions 
imply that in the case when the ratio of specific heats $\gamma>1$, the velocity difference across 
zones of the shock is smaller than in the case when $\gamma=1$.  This may qualitatively explain 
the steeper PDF of convergence in molecular turbulence.

The PDF of the velocity itself is not very well understood \citep{Elmegreen04}.
The usual argument is that the velocity PDF should be Gaussian is based on the 
central-limit theorem applied to Fourier components of the velocity field.
This argument, however, does not take into account correlations which are
important for energy transfer among scales. In fact, the velocity PDF must possess
non-zero skewness for the transfer to be possible \cite[see, e.g.,][]{Lesieur97}.

Analytical studies of 
Burgers' (shock-dominated) turbulence found that the gradient of the velocity 
has power law asymptotics (same as density, see \cite{Frisch01}).  
We do not see this effect in our simulations, perhaps because of our
limited numerical resolution (see also \S\ref{ssec:pdf}).

\begin{figure}
\centering\includegraphics[width=0.8\linewidth]{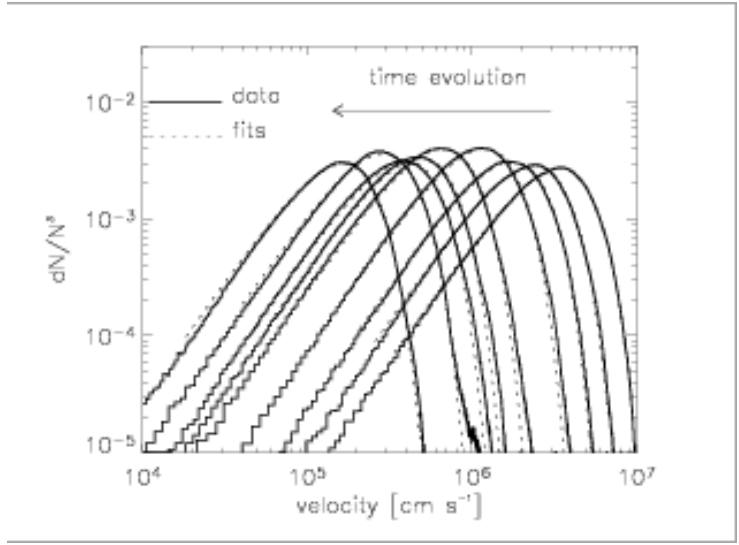}
\caption[Maxwellian distribution of velocity]%
{PDFs of the velocity magnitude during the \molt\ simulation. The peak
shifts toward smaller values as the turbulence decays.  Plotted curves correspond
to the same moments of time as curves on Figs.~\ref{vels:fig2} \& \ref{vels:fig3}.  
The arrow points in the direction of time (velocities decrease).  Dotted lines show Maxwellian
fits.}
\label{vels:fig4}
\end{figure}

In our simulations, the initial distribution of the velocities has imposed Gaussian 
statistics. We find that deviations from the Gaussian distribution do occur but they are 
minor. To illustrate this, in Fig.~\ref{vels:fig4} we plot the PDFs of the velocity magnitude, 
$v = \sqrt{x^2+y^2+z^2}$, which takes the form of a Maxwell distribution, 
i.e. $\exp\{-A v^2\}v^2$, as expected.

\section{SIMULATED OBSERVATIONS} 
\label{sec:obs}

\subsection{Background} 
\label{ssec:bac}

Observational predictions from numerical results have generally involved the post-processing 
of isothermal flows to generate simulated observations \citep{Falgarone94,Padoan98,Lazarian01,Ossenkopf02,
Brunt03}.  We can validate such an approach by comparing synthetic maps from 
isothermal models to similar maps from the molecular models.
We note that the compact and dense regime taken in our simulations is not directly comparable to 
existing observations of molecular clouds but rather corresponds to dense cores. 

Interstellar clouds appear to follow a scaling relation first inferred
from an analysis of observational data by \citet{Larson81}. Larson used 
\ch{^{13}CO} (optically thin line) maps of different molecular clouds
to establish a size--linewidth relation
\begin{equation}
\label{obs:eq2}
\sigma_v \propto \lambda^{b},\quad b=0.5\pm0.2,
\end{equation}
where $\sigma_v$ is the velocity dispersion and 
$\lambda$ is the size of the region.

A theoretical explanation for this scaling relation can be
inferred from the following argument.  The velocity dispersion, $\sigma_v$, can be
approximated by the square root of the second order velocity structure function,
\begin{equation}
\label{obs:eq3}
   S_2 = \avrg{\left(\bm{v}(\bm{r}) - \bm{v}(\bm{r}')\right)^2}\propto
\lambda^{\zeta_2},\quad\lambda\equiv|\bm{r} - \bm{r}'|.
\end{equation}
The value of $\zeta_2$ is determined from the model of driven supersonic
turbulence proposed by \citet{Boldyrev02a}, based on the dimensionality of the 
dissipative structures.  Boldyrev's theory suggests the following scaling law:
\begin{equation}
\label{obs:eq4}
   \sigma_v = \lambda^{0.37},
\end{equation}
which is in good agreement with the above cited observational value.  Pure
K41 scaling gives smaller values of the exponent ($b\approx0.33$). However, in the case
of decaying turbulence Boldyrev's scaling laws do not apply.  As we have 
demonstrated in Section~\ref{sec:eng}, decaying turbulence results in
a shallower power law energy spectrum  than follows from K41 theory.  
To estimate the exponent $b$ in this case 
we can use the connection between velocity and power spectrum statistics,
\begin{gather}
\label{obs:eq5}
   P_1(k) \propto k^{-m},\\
\label{obs:eq6}
   S_2 \propto \lambda^{m-1},
\end{gather}
where $P_1(k)$ is the (one dimensional) power spectrum; using the notation of 
section~\ref{sec:eng}, $n = m+2$  \citep[see, e.g.,][]{Frish95}.  Using 
values of $n$ given in Table~\ref{eng:tab1}, $m\approx-0.88\ldots-1.16$,
which implies, $b=-0.06\ldots0.08$.  Hence, we can predict that $\sigma_v$
should not exhibit strong scaling in the case of decaying turbulence.

Deduction of Larson's scaling law from the observations is not a straightforward task.  
The width of a spectral line, broadened by the turbulent motions, is essentially the 
width of a composite line, contaminated by all the gas motions along the line of sight. How
the width of such a composite line should depend on the size of the sampled region is not
obvious.  If we sample $\sigma_v$ from the regions that appear to be cores (i.e. continuous 
prominent patches of radiation intensity), sorting the $\sigma_v$ measurement by the size 
of such cores, we might stand a better chance of recovering true velocity scaling.  
This is because such cores may actually be coherent 3D structures, although there is no
way we can protect our statistics from unavoidable mistakes \citep{Vazquez97,Pichardo00,
Ballesteros02}.  We have no means of 
distinguishing between ``false'' cores --- column density enhancements produced by 
overlapping, unrelated density enhancements assembled by chance along the line of sight --- 
and true cores --- compact 3D objects. In regions of active star formation where 
self-gravity effects are important, we can hope that the number of such mis-identifications 
will be effectively minimised.

In the case we are studying, there is no boosting of contrast due to self-gravity but there 
are additional contrast parameters: the chemical composition and temperature distributions.  
Indeed, our simulations provide detailed pictures of the \ch{H_2} and temperature distributions.  
Non-uniform chemical tracers may provide additional contrasts to the maps. What sort of regions 
can chemical tracers emphasise?  As we found out in Paper~I, the distribution of molecules in 
decaying turbulence is rather homogeneous, and there is no evidence that the \ch{H_2} 
distribution alone will highlight 3D structures within the simulated region. However, 
the contrast in temperatures is far more distinct. Most of the gas in the simulated domain stays 
cool due to the efficient cooling; only sufficiently strong shocks can temporarily heat up the gas 
in certain regions, up to several thousand degrees. Hence, if we select a tracer 
sensitive to the temperature, we would be able to construct maps of regions with 
different values of temperature.

Temperature sensitive tracer maps may help answer another question 
relevant to the present study. Is detailed information about molecule and temperature distributions 
crucial for predicting emission, or is the zero-order approximation of isothermality sufficient?

\subsection{Scaled projection of CO emission} 
\label{sec:scp}

\begin{figure}
\centering
\begin{minipage}[b]{\linewidth}
\centering\includegraphics[width=.6\linewidth]{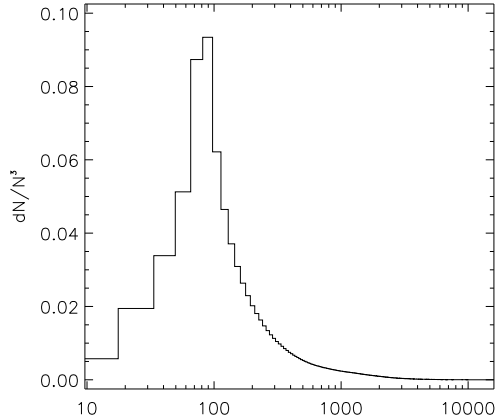}
\caption[Histogram of the temperature distribution at $t=60$~yr]%
{Histogram of the temperature distribution at $t=60$~yr. The
histogram has $1000$ bins with size $\delta T = 16$~K.  A large fraction of the
gas has temperature close to $T=100$~K.  The average temperature is $\approx630$~K.
Data taken from the \molt\ simulation.}
\label{fig:scp2}
\end{minipage}
\end{figure}

The gas in our simulation is relatively warm with an average temperature $\gtrsim100$~K 
(Fig.~\ref{fig:scp2}). Rotational emission  lines of \ch{^{12}CO} are a suitable tracer 
in this temperature range \citep{McKee82}.  Emission from the transitions  
between states with different rotational quantum number $J$ (only 
$J\rightarrow J-1$ transitions are allowed due to the quantum selection
rule) will highlight regions with different temperatures. The choice
of this particular tracer is reasonable because our molecular code actually
computes the equilibrium  abundance of \ch{CO} to account for the cooling through 
rotational and vibrational \ch{CO} emission.

We take the non-LTE formulae presented by \citet{McKee82}. We assume that each 
emission line is optically thin and, therefore, the intensity of the 
emission (the ``brightness'' of the pixel on the emission map) is 
proportional to the column density of \ch{CO}, 
\begin{equation}
\label{equ:coe1}
   \text{emission at $J$ level }\propto \sum \avrg{n_J(\ch{CO})}_i \Delta h_i
\end{equation}
where $\avrg{n_J(\ch{CO})}_i$ is the average number density of \ch{CO} 
at level $J$ in the cell of size $(\Delta h_i)^3$.

The simplest way to construct a synthetic map of \ch{CO} emission is to map the 
column density, as prescribed by equation~(\ref{equ:coe1}).  
Following the work of \citet{Ostriker01}, we identify a clump as
a region of contrast (ROC), which is regarded as such if its \ch{CO} 
column density is at least a factor of $c_r$ larger than the mean column
density of \ch{CO} for the entire map. This procedure is somewhat similar
to the procedure of subtracting ``background'' in the reduction of observational 
data.  In our current study, we fix $c_r=1$, i.e. all regions with column density 
larger than the mean column density are identified as ROCs, and we constrain our 
analysis to the data we infer from the ROCs only.

To deduce the  relationship between line width and size from the synthetic maps,  
we need an algorithm for the detection of ROCs of different sizes. One means of 
achieving this is by constructing ``blurred'' maps of lower resolution by 
rescaling the original map of \ch{CO} column density by a certain 
factor, as proposed by \citet{Ostriker01}.  We rescale the original $256^2$
grid by a factor of $2^s$, where $s=0,1,2,\ldots,7$.  This results in
eight maps with resolutions $2^2,\,4^2,\,8^2,\ldots,\,256^2$, which 
we can employ for ROC identification.  

To construct \ch{CO} emission maps from the \isot\ simulation with $T=100$~K we need to 
determine an appropriate abundance of \ch{CO} (in each computational zone) 
without having information about the \ch{H_2} abundance. The chemistry algorithm 
implemented in our version of ZEUS-3D implies that, for this temperature 
all free gas-phase carbon is in the form of \ch{CO} for all plausible 
mass densities and \ch{H_2} fractions. The total abundance of \ch{C} is fixed in 
the code, $\mathfrak{f}(\ch{C}) = 2.0\times10^{-4}$.  Hence, in the isothermal 
case we take the distribution of \ch{CO} as homogeneous with the fixed value 
of $2.0\times10^{-4}$. 

We select the time when the largest fraction of the 
molecular gas is at $T=100$~K (see the temperature histogram in Fig.~\ref{fig:scp2}).
The actual shapes of lines and maxima in \ch{CO} emission will depend on the 
underlying mass density distribution and velocity along the line of sight.  

To be able to compare the morphological features directly 
we use density and velocity data from the \molt\
simulation in {\em both cases}. The actual distributions of \ch{CO} and 
temperature are employed for the \molt\ maps, and homogeneous distributions 
of \ch{CO} and temperature for the \isot\ maps. We will refer to such 
isothermal data as {\em quasi-isothermal}. As we will show, the results of the analysis 
of the quasi-isothermal maps are very similar to the results of the analysis with the 
actual isothermal simulation data.  This is not surprising giving that the statistical 
properties of the density and velocity fields in both simulations are very similar, as 
established in \S\S~\ref{ssec:vels} and \ref{ssec:dens}.

For each ROC, we compute the mass-weighted dispersion of the line-of-sight velocity 
$\sigma_v$, which represents the line width for a region of projected area 
$[(L/N)/2^s]^2$. We record $M(\ch{CO_J})$ within each ROC, and the
virial parameter,
\begin{equation}
\label{equ:scp1}
   \alpha=\frac{5\sigma_v^2[(L/N)/2^s]}{GM(\ch{CO_J})},
\end{equation}
where $M(\ch{CO_J})$ is the mass of the \ch{CO_J} (\ch{CO} responsible for
 $J\rightarrow J-1$ emission) and $G$ is Newton's gravitational constant 
\citep{Ostriker01}. The parameter $\alpha$ is directly proportional to the 
ratio of kinetic energy to gravitational energy, and represents a measure of the
relative importance of gravity \citep{McKee92,Ostriker01}.

\subsection{Analysis of the emission maps} 
\label{ssec:aem}

We now present the results of the synthetic observations.  
For the data sets discussed in section~\ref{sec:scp}, we have constructed 
\ch{CO}~(4--3) emission maps that highlight the gas at 
temperatures around $100$~K (for gas with average density of 
$10^6$\cm{-3}), and \ch{CO}~(20--19), which traces much 
hotter gas of about $1000$~K \citep{McKee82}.

\begin{figure}
\centering\includegraphics[width=0.95\textwidth]{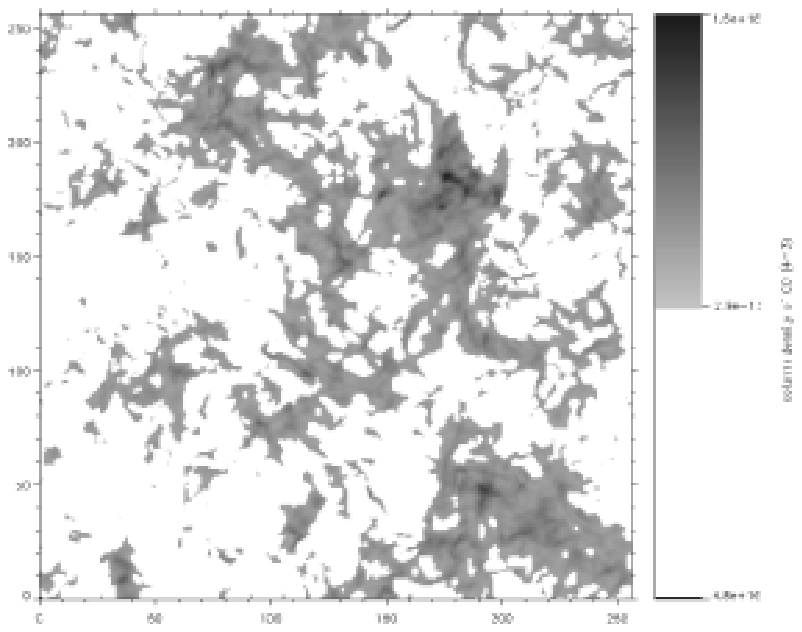}
\centering\includegraphics[width=0.95\textwidth]{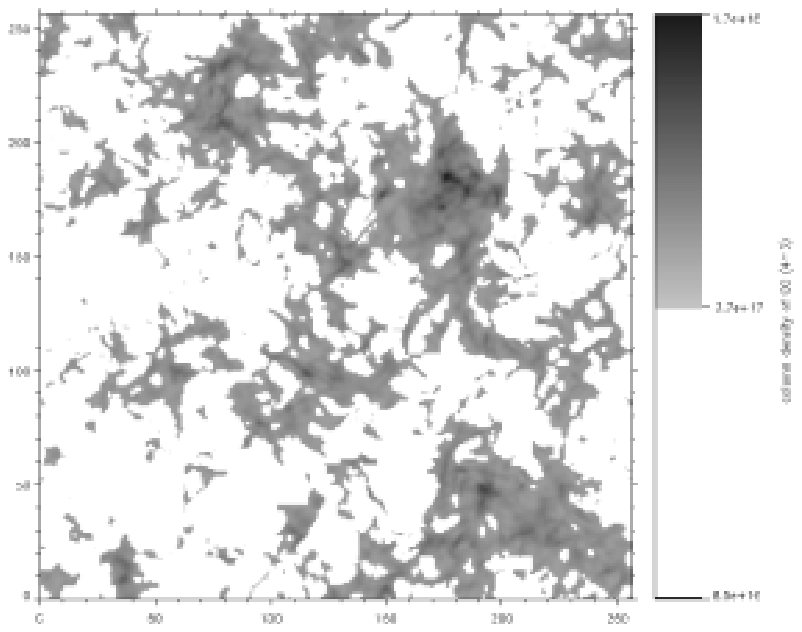}
\caption{Maps of  \ch{^{12}CO} $J=4\rightarrow3$ emission constructed from the \molt\ 
data (upper panel), and the same emission map constructed from the quasi-isothermal 
turbulence data (lower panel), as described in section~\ref{sec:scp}.} 
\label{fig:aem1}
\end{figure}

\ch{CO}~(4--3) emission maps for the \molt\ data and 
the quasi-isothermal data are presented in Fig.~\ref{fig:aem1}. The similarity of these 
maps is quite remarkable. Differences are difficult to detect.  
This supports the claim made in Paper~I that isothermal simulations are
suitable for mapping {\em molecular} emission.

\begin{figure}
\begin{minipage}[b]{.45\linewidth}
\centering\includegraphics[width=\linewidth]{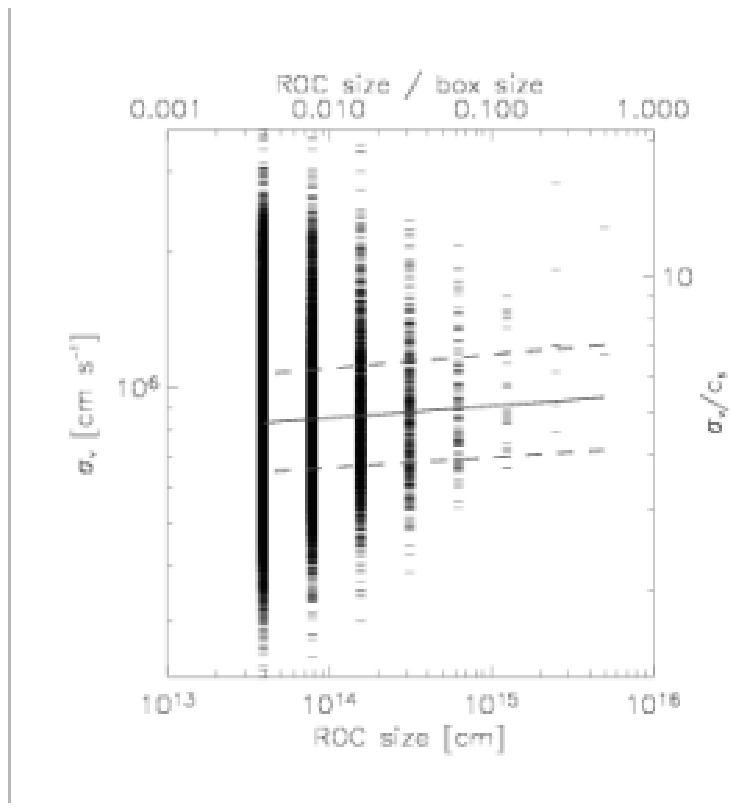}
\caption[Sigma vs. size of ROC for $J=4\rightarrow3$ emission map]%
{Sigma vs. size of ROC for the $J=4\rightarrow3$ molecular turbulence
map. For reference, $c_s$ is the mass-weighted average speed of sound.}
\label{fig:aem2}
\end{minipage}\hfill
\begin{minipage}[b]{.45\linewidth}
\centering\includegraphics[width=\linewidth]{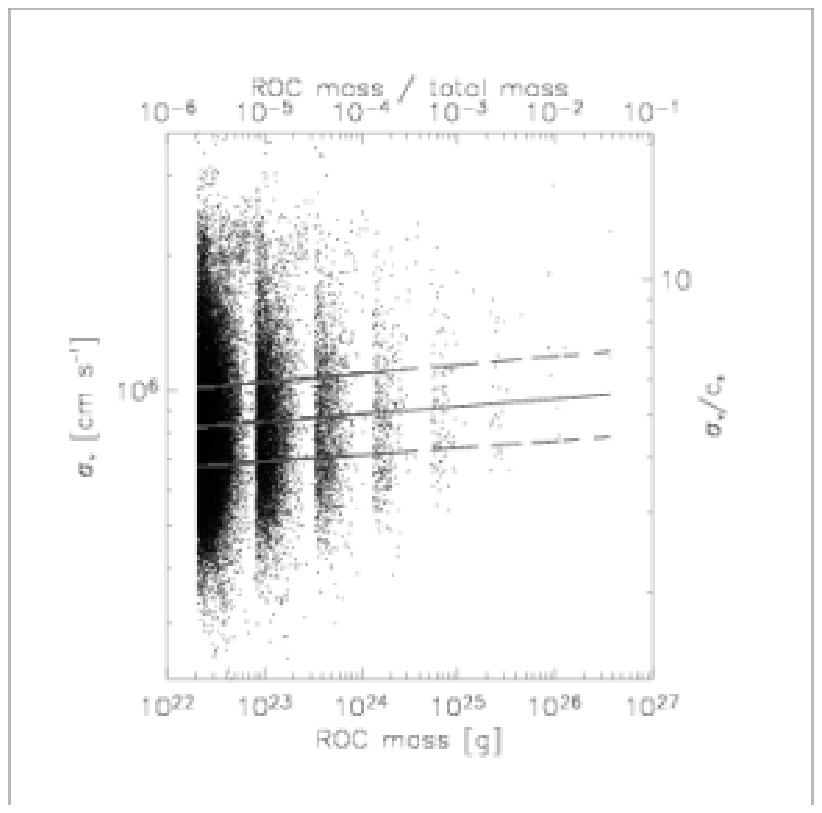}
\caption[Sigma vs. mass of ROC, for $J=4\rightarrow3$ emission map]%
{Sigma vs. mass of ROC, for $J=4\rightarrow3$ molecular turbulence
map. For reference, $c_s$ is the mass-weighted average speed of sound.}
\label{fig:aem3}
\end{minipage}\\
\begin{minipage}[b]{.45\linewidth}
\centering\includegraphics[width=\linewidth]{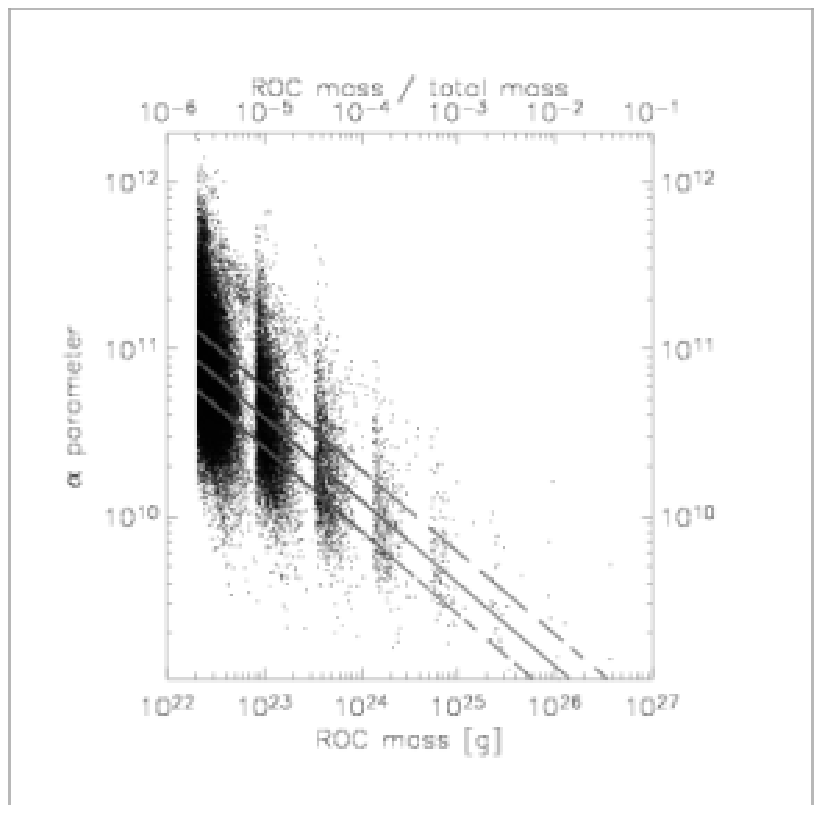}
\caption[Alpha parameter vs. mass of ROC for $J=4\rightarrow3$ emission map]%
{Alpha parameter vs. mass of ROC for $J=4\rightarrow3$ 
molecular turbulence map}
\label{fig:aem4}
\end{minipage}\hfill
\begin{minipage}[b]{.45\linewidth}
\centering\includegraphics[width=\linewidth]{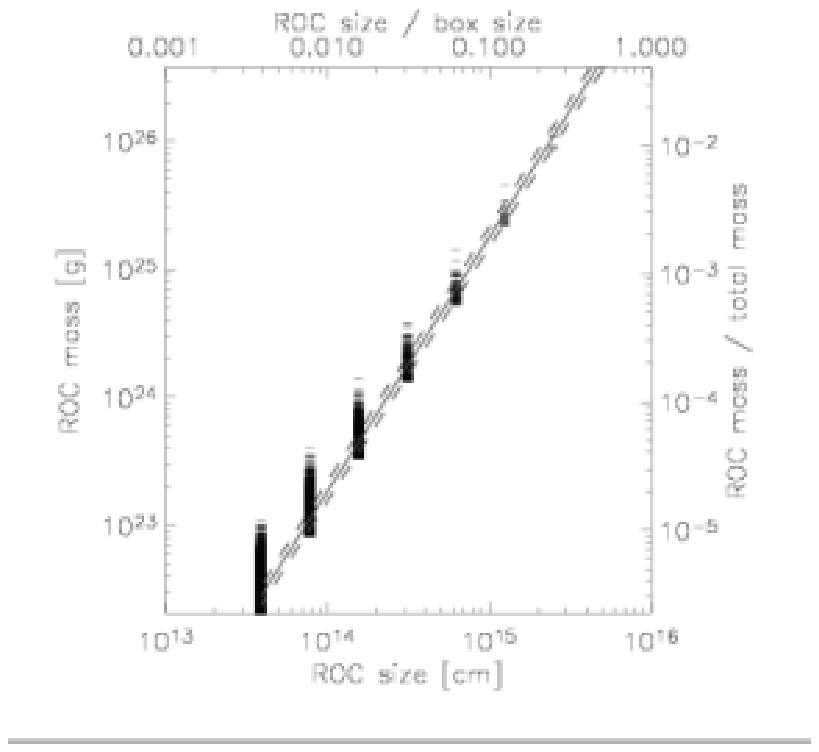}
\caption[Mass of ROC vs. size of ROC, for $J=4\rightarrow3$ emission map]%
{Mass of ROC vs. size of ROC, for $J=4\rightarrow3$ 
molecular turbulence map}
\label{fig:aem5}
\end{minipage}
\end{figure}

An analysis of the statistical properties of these maps also reveals no significant 
discrepancies. In Figs.~\ref{fig:aem2}--\ref{fig:aem5}, we display scatter plots of 
the parameters of the ROCs which were identified on the \molt\ maps at eight different 
resolutions. Each point or small dash on the plots corresponds to the parameters of 
a ROC.  Although there is a general trend in mean dependences, there is plenty of 
scatter in all the plots. The solid lines represent least-square linear fits of the form
\begin{equation}
\label{equ:aem1}
   \frac{\text{d}\log\sigma_v(\lambda)}{\text{d}\log \lambda} = b,
\end{equation}
where $\lambda=(L/N)/2^s$ is the size of the ROC;
\begin{equation}
\label{equ:aem2}
   \frac{\text{d}\log\sigma_v(\lambda)}{\text{d}\log M_\lambda(\ch{CO_J})} = b';
\end{equation}
\begin{equation}
\label{equ:aem3}
   \frac{\text{d}\log\alpha}{\text{d}\log M_\lambda(\ch{CO_J})} = c;
\end{equation}
\begin{equation}
\label{equ:aem4}
   \frac{\text{d}\log M_\lambda(\ch{CO_J})}{\text{d}\log \lambda} = d.
\end{equation}
The parallel dashed lines in the figures mark the $1\sigma$ deviations 
for the fits.  The values (\ref{equ:aem1}) -- (\ref{equ:aem4}) 
deduced from the \molt\ data are as follows,

\begin{equation}
\label{equ:aem5}
\begin{split}
   b  &= (2.6\pm0.4)\times10^{-2},\quad
   b' = (1.7\pm0.2)\times10^{-2},\\
   c  &= -(4.94\pm0.04)\times10^{-1},\quad
   d  = 1.984\pm0.003.
\end{split}
\end{equation}
The values $b$, $b'$, $c$ and $d$ deduced from the map constructed from 
the quasi-isothermal data are the same within the error estimates:
\begin{equation}
\label{equ:aem6}
\begin{split}
   b  &= (2.7\pm0.4)\times10^{-2},\quad
   b' = (1.7\pm0.2)\times10^{-2},\\
   c  &= -(4.86\pm0.04)\times10^{-1},\quad
   d  = 1.985\pm0.003.
\end{split}
\end{equation}

As noted in \S~\ref{sec:scp}, the statistical properties of 
the emission map constructed using density and velocity distributions
from the actual \isot\ simulations are virtually the same.  The value of 
parameters in that case are
\begin{equation}
\label{equ:aem7}
\begin{split}
   b  &= (1.7\pm0.4)\times10^{-2},\quad
   b' = (1.1\pm0.2)\times10^{-2},\\
   c  &= -(4.99\pm0.04)\times10^{-1},\quad
   d  = 1.985\pm0.002.
\end{split}
\end{equation}

Such a good agreement in the morphology and statistical properties of the maps
is due to the very homogeneous distribution of molecules in the gas. At this stage 
of molecular turbulence evolution, $t=60$~yr, there are numerous dissociative shocks, 
and the mean characteristics would suggest that molecular turbulence should be very 
different from isothermal turbulence in many respects (at $t=60$~yr the average 
molecular hydrogen fraction is $\avrg{f}=0.1$, and average temperature is 
$\avrg{T}=630$~K). Despite this, the analysis of actual molecular emission maps
provides evidence to the contrary.

Furthermore, the shapes of spectral lines in both cases are very similar.  
The line shapes are very roughly Gaussian, with exponential tails in many cases.  
Exponential tails have indeed been observed and can be theoretically explained by the 
intermittency of the velocity distribution \citep{Falgarone94}.

The linewidth-size relationship found for the ROCs is very flat and does not
correspond to that of Larson's law described in \S~\ref{ssec:bac}. This has been fully 
discussed by \citet{Ostriker01}, who determined that the least squares fit is associated 
with the superposition of density structure along the line of sight. The relation for 
actual clumps or cores is related to the lower envelope of the linewidth-size scatter plot,
which has a slope of $\sim0.3\pm0.1$, toward the shallow end of the observed values given 
by equation~\ref{obs:eq2}.

\begin{figure}
\centering\includegraphics[width=0.95\textwidth]{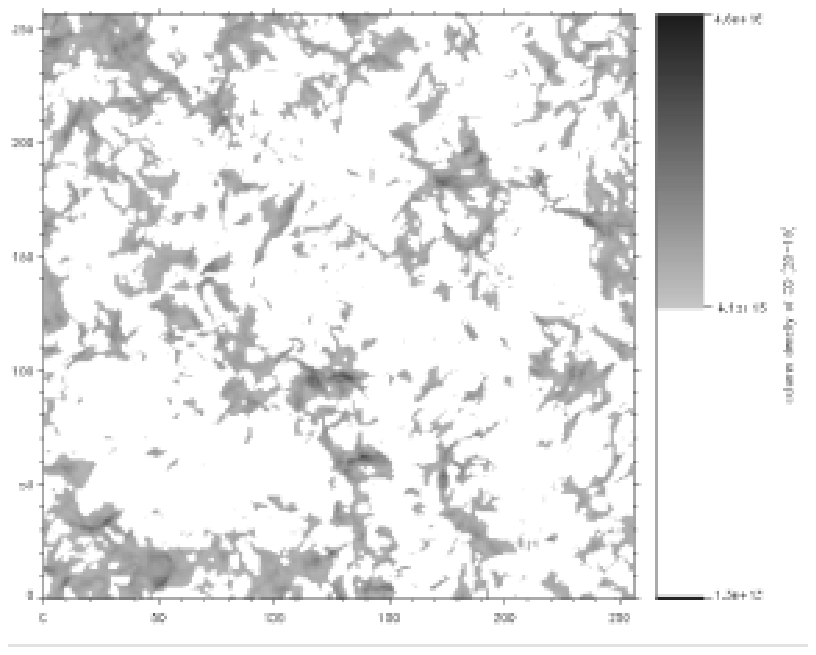}
\centering\includegraphics[width=0.95\textwidth]{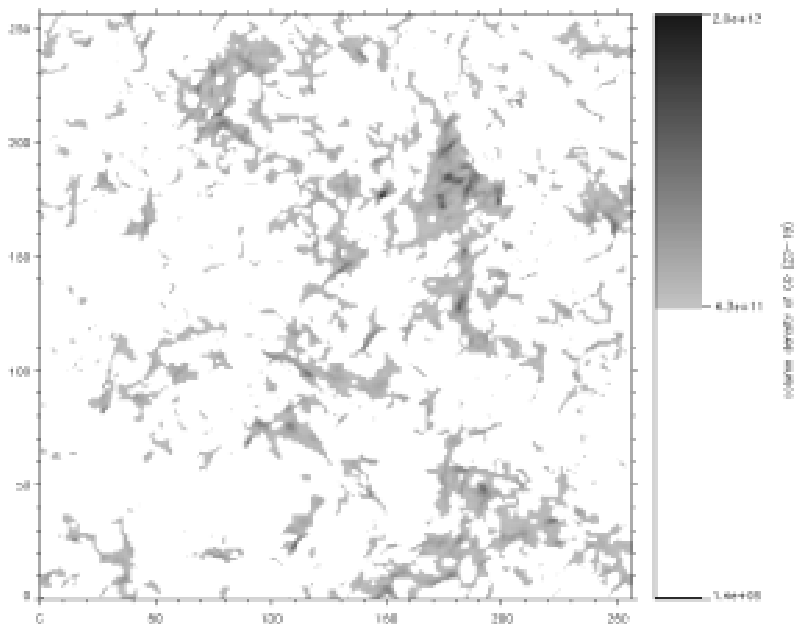}
\caption{Maps of \ch{^{12}CO} $J=20\rightarrow19$ emission constructed 
from the \molt\ data (upper panel). The same emission map constructed from the 
quasi-isothermal turbulence data (lower panel) as described in
\S~\ref{sec:scp}. Note the huge difference in intensity as given by the colour bars. } 
\label{fig:aem6}
\end{figure}

A similar analysis of synthetic \ch{CO_J}(20--19) maps 
emphasises the differences between the isothermal and non-isothermal regimes.  
Emission maps in this \ch{CO} line are drastically different, as displayed in 
Fig.~\ref{fig:aem6}. The peak intensity in the isothermal case is a factor of more
than $10^3$ lower.  This is to be expected, as the $J=20\rightarrow19$ transition 
traces only those regions where the gas has $T\approx1000$~K.
In the case of \molt, hot regions are naturally associated with strong shocks, whereas 
in the (quasi-) isothermal case the temperature is much lower and uniform. Therefore, 
in the isothermal case, the emission in this 
line simply reflects the column density (compare the lower
maps of Figs.~\ref{fig:aem1} and \ref{fig:aem6}). 

For the $J=20\rightarrow19$ emission maps, the values of parameters 
(\ref{equ:aem1}) --- (\ref{equ:aem4}) are again given by least-square fits.
Emission maps based on the \molt\ simulation data yield
\begin{equation}
\label{equ:aem8}
\begin{split}
   b  &= (4.7\pm0.4)\times10^{-2},\quad
   b' = (2.7\pm0.2)\times10^{-2},\\
   c  &= -(5.01\pm0.04)\times10^{-1},\quad
   d  = 1.950\pm0.002.
\end{split}
\end{equation}
and for the quasi-isothermal case,
\begin{equation}
\label{equ:aem9}
\begin{split}
   b  &= (6.2\pm0.7)\times10^{-2},\quad
   b' = (2.7\pm0.3)\times10^{-2},\\
   c  &= -(5.71\pm0.06)\times10^{-1},\quad
   d  = 1.936\pm0.007.
\end{split}
\end{equation}.
It is remarkable that statistical properties of the maps of \ch{CO}(20--19) as summarised by
equations (\ref{equ:aem8}) and (\ref{equ:aem9}) do not reveal any differences between 
\molt\ and \isot, although the underlying maps are completely different.  They are very similar
to the parameters of \ch{CO}(4--3) emission, see Eqs.~(\ref{equ:aem5}) -- (\ref{equ:aem7}).  Linewidth-size
relationship is somewhat steeper for the \ch{CO}(20--19) maps, which might be related to the size of
the statistical sample (number of ROCs is smaller for $J=20\rightarrow19$ transition).

The $J=20\rightarrow19$ emission line profiles derived from molecular data are 
somewhat different from the line profiles derived from the quasi-isothermal data, 
with the emission lines often more prominently peaked (though far weaker) 
in the quasi-isothermal case.

The remaining parameters are qualitatively similar to the parameters
derived for the $J=4\rightarrow3$ line emission maps.  The virial parameter $\alpha$ is
large ($\alpha>10^{10}$), confirming that the kinetic energy is the dominant 
form of energy in this turbulence regime, and the `cores' on the maps 
(ROCs) are false cores which cannot be gravitationally confined.  
The fact that $c<0$ indicates that gravity is relatively more 
important for regions with small mass (i.e. of smaller size).  

The scale dependence of ROC mass and the virial parameter $\alpha$ have 
the following interpretation. In the limit of constant column density,
the mass of a ROC of size $\lambda$ is $M\propto\rho_\text{col}\lambda^2$, 
where $\rho_\text{col}$ is the column density, and the resulting scale dependence
is $M\propto\lambda^2$. These are close to the values we deduce from all
emission maps, $M\propto\lambda^d$, where $d=1.96\pm0.03$, as concluded 
also by \citet{Ballesteros02}.
Scaling of the virial parameter $\alpha$ follows from the result 
for the mass scaling.  Given that $\sigma$ is independent of 
$\lambda$, the scaling dependence of the parameter $\alpha$ approaches 
$\propto M^{-0.5}$ (in the limit of constant column densities).  
The numerical value is close to this limit: we find $\alpha\propto M^c$, 
where $c=-0.5\pm0.2$.

\subsection{Discussion} 
\label{sec:dis}

For molecular line emission that traces the bulk of the gas,
isothermal simulations can be very accurate in mapping molecular emission.
Properties of the emission maps and shapes of emission lines reproduce 
the corresponding values deduced from the detailed molecular simulations.
An attempt to map shocked regions with a temperature-sensitive chemical
tracer fails in the case of isothermal turbulence, as is naturally anticipated.  
Results of isothermal turbulence simulations should only be used
for what they have been made --- tracing gas close to the average temperature.

The predictions of values of 
scaling exponents from analysis of emission maps with different CO rotational 
number $J$ are approximately equal.

The algorithm for construction of synthetic emission maps used here is very simple.  
It functions only for the case of optically thin emission and may serve as a starting 
point for a deeper analysis.  Self-consistent three-dimensional radiative transfer  
algorithms are essential for obtaining reliable predictions as, for example, used by 
\citet{Padoan98} based on the radiative transfer code of \citet{Juvela97}.  The large 
velocity gradient approximation \citep{Sobolev57} for radiative transfer was used by 
\citet{Ossenkopf02} to construct synthetic emission maps.

\section{CONCLUSIONS} 
\label{sec:con}
 
We have examined how molecular dynamics influences
supersonic turbulence and, {\em vice versa}, how turbulence influences
molecular content. Numerical simulations of cloud dynamics have usually employed an isothermal
equation of state as an approximation for 
gas inside molecular clouds. Hence, we have compared our results
(\molt) to that of the equivalent isothermal 
simulation (\isot). We analysed decaying turbulence, beginning from a molecular state in which
strong shocks develop and destroy 87\% of the molecules and running until 80\% of the 
molecules have reformed. Our main results are as follows.
\begin{enumerate}

\item There is no significant difference between the evolution of power spectra for  \molt\ and \isot. 
The spectra are not power laws but involve a faster decay in the energy associated with the higher 
wavenumber regime.

\item Despite strongly supersonic root-mean-square velocities, the ratio of energy in 
compressional modes to energy in solenoidal modes at low wavenumbers drops from 
the initial value of 1:2 to $\sim$1:5 almost immediately.  Little further 
decrease occurs at the low wavenumber end while the flow remains supersonic.

\item Compression waves steepen much faster in the simulation of molecular turbulence leading to 
a rapid increase in high wavenumber compressional energy. This appears to be due to the slower 
dissipation of thermal energy during the dissociative phase than in the equivalent isothermal phase. 
In contrast, the high wavenumber compressional energy in \isot\ later 
overtakes that of the \molt.

\item The effective polytropic index in  \molt\ varies considerably. In the very early stages, 
there are effectively two isothermal phases, corresponding to atomic gas at $\sim$ 8000\,K and 
molecular gas at  $\sim$ 40\,K.  The index is then reduced to below and above unity for extended 
periods (see Fig.~\ref{dens:fig2}) before approaching unity at late times. The sub-isothermal period 
is related to the presence of strong compression and molecule reformation behind fast shocks.

\item The density PDFs in the \molt\ case are found to be 
close to log-normal. The strongest deviations occur early on when strong shocks distort the statistics.
The deviations are smaller in the \molt\ case than in \isot, probably due to the higher effective 
polytropic index. 
 
\item The structure of the velocity fields was studied with the help of the
divergence and modulus of curl. Vortex isosurfaces take the form of sheets closely related in 
space and size to the shock surfaces. 

\item The PDFs of vorticity, as well as velocity gradients, take the form of exponentials.  
We constructed  PDFs of {\em compensated} distributions where the change of the mean values of the
distributions is accounted for in the exponentials. With
constant parameters, the functions provide a useful tool for diagnosing
decaying turbulence. Velocity PDFs maintain Gaussian distributions. 

\item Simulated maps of CO rotational emission lines are constructed. We find that isothermal simulations
provide excellent agreement with the molecular predictions for tracers of the cool bulk of the gas. This is
confirmed through the `Region of Contrast' method to characterise the properties of the emission maps. 
In Paper~I,  we showed that as turbulence decays, molecules do not remain within swept-up shells 
but are distributed very homogeneously, and the molecular fraction distribution has a small variation.  
This results in very similar looking mass and molecular mass distributions during the
reformation phase.  On the basis of this result it is argued that
isothermal simulations can be used to map molecules.  This hypothesis is
supported by the synthetic maps. 

\item On the other hand, large differences are found in the emission distribution of high-J 
CO lines. These lines are sensitive to the temperature. 

\item  In astrophysical turbulence, the magnetic field 
influences
turbulent cascades \citep{Vestuto03}.  For example, the
column density power spectrum was found to be  significantly 
shallower as the 
field strength is raised \citep{Padoan04}. A strong magnetic 
field also enhances the shock number transverse to the field direction at the 
expense of parallel shocks \citep{Smith00b}. 

The magnetic field would also weaken shocks, allowing molecules to survive 
the passage of stronger shocks. This will tend to reduce the differences with 
respect to the \isot\ case caused by the chemical and cooling processes. 

\end{enumerate}

The main conclusion of this work is that isothermal simulations adequately
model molecular turbulence and can be used to predict many properties
of the molecular emission; given the supersonic dynamics, the chemical
reactions in a turbulent gas can be significantly accelerated due to
strong compression and advection.

This work provides a basis for future research in a number of directions.
It is possible that the behaviour of molecules can be modelled as a passive 
scalar, and pseudo-temperature distributions can be constructed from 
appropriately gauged divergence fields.

\section{Acknowledgements} 
We thank the anonymous referee for a careful report, and in particular
for pointing out the work on PDF asymptotics.
The computations reported here were performed using the UK
Astrophysical Fluids Facility (UKAFF) and FORGE (Armagh),
funded by the PPARC JREI scheme, in collaboration with SGI. M-MML
was partially funded by the NASA Astrophysical Theory Program
under grant number NAG5-10103 and NSF grants AST99-85392, and AST03-07793. 
GB has been partially funded by the PPARC.  Armagh Observatory receives
funding from the Northern Ireland Department of Culture, Arts and Leisure.
ZEUS-3D was used by courtesy of the Laboratory of Computational 
Astrophysics at UC San Diego.  This research has made use of NASA's Astrophysics 
Data System Bibliographic Services.

\bibliography{gbp}

\end{document}